\documentclass[sigconf]{acmart}

\AtBeginDocument{%
  \providecommand\BibTeX{{%
    \normalfont B\kern-0.5em{\scshape i\kern-0.25em b}\kern-0.8em\TeX}}}

\usepackage[T1]{fontenc}
\usepackage[utf8]{inputenc}
\usepackage{listings}
\usepackage{multirow}
\usepackage{comment}
\usepackage{subfig}
\usepackage[export]{adjustbox}
\usepackage{enumitem}
\usepackage{amsmath}
\mathchardef\mhyphen="2D 

\newcommand{\finetuning}{\textit{fine-tuning}}
\newcommand{\catforget}{\textit{catastrophic forgetting}}
\newcommand{\kndistill}{\textit{knowledge distillation}}

\DeclareMathOperator*{\argmin}{arg\,min}

\setcopyright{none}
\renewcommand\footnotetextcopyrightpermission[1]{} 
\usepackage{fancyhdr}
\fancypagestyle{mystyle}{%
    \fancyhead[R]{Accepted as a conference paper for SIGMOD 2023 @Seattle, WA, USA}\fancyhead[L]{ }
}%
\begin{document}

\title{Detect, Distill and Update: \\ 
Learned DB Systems Facing Out of Distribution Data}

\author{Meghdad Kurmanji , Peter Triantafillou}
\affiliation{\institution{University of Warwick, Coventry, UK}\country{}}
\email{{meghdad.kurmanji, p.triantafillou}@warwick.ac.uk}

\begin{abstract}
Machine Learning (ML) is changing DBs as many DB components are being replaced by ML models. One open problem in this setting is how to update such ML models in the presence of data updates. We start this investigation focusing on data insertions (dominating updates in analytical DBs). We study how to update neural network (NN) models when new data follows a different distribution (a.k.a. it is "out-of-distribution" -- OOD), rendering previously-trained NNs inaccurate. A requirement in our problem setting is that learned DB components should ensure high accuracy for tasks on old and new data (e.g., for approximate query processing (AQP), cardinality estimation (CE), synthetic data generation (DG), etc.).

This paper proposes a novel updatability framework (DDUp). DDUp can provide updatability for different learned DB system components, even based on different NNs, without the high costs to retrain the NNs from scratch. DDUp entails two components: First, a novel, efficient, and principled statistical-testing approach to detect OOD data. Second, a novel model updating approach, grounded on the principles of transfer learning with knowledge distillation, to update learned models efficiently, while still ensuring high accuracy. We develop and showcase DDUp's applicability for three different learned DB components, AQP, CE, and DG, each employing a different type of NN. Detailed  experimental evaluation using real and benchmark datasets for AQP, CE, and DG detail DDUp's performance advantages.

\end{abstract}


\maketitle
\thispagestyle{mystyle}
\pagestyle{plain}

\section{Introduction} \label{introduction}
Database systems (DBs) are largely embracing ML. With data volumes reaching unprecedented levels, ML can provide highly-accurate methods to perform central data management tasks more efficiently. Applications abound: AQP engines are leveraging ML to answer queries much faster and more accurately than traditional DBs \cite{ma2019dbest,hilprecht2019deepdb,thirumuruganathan2020approximate,ma2021learned}.
Cardinality/selectivity estimation, has improved considerably leveraging ML  \cite{yang2019deep,yang2020neurocard,hasan2020deep,zhu2020flat,wang2020we}. Likewise for query optimization 
\cite{marcus2019neo,kipf2018learned,marcus2021bao},
indexes \cite{kraska2018case,ding2020alex,nathan2020learning,ding2020tsunami}, cost estimation \cite{zhi2021efficient, siddiqui2020cost}, workload forecasting \cite{zhu2019novel}, DB tuning \cite{van2017automatic,li2019qtune,zhang2019end}, synthetic data generation \citep{xu2019modeling,choi2017generating,park2018data}, etc. 

\subsection{Challenges}\label{challenges}
As research in learned DB systems
matures, two key pitfalls are emerging. First, if the "context" (such as the data, the DB system, and/or the workload) changes, previously trained models are no longer accurate. Second, training accurate ML models is costly. Hence, retraining from scratch when the context changes should be avoided whenever possible.
Emerging ML paradigms, such as active learning, transfer learning, meta-learning, and zero/few-shot learning are a good fit for such context changes and have been the focus of recent related works   \cite{ma2020active, hilprecht2021one, wu2021unified}, where the primary focus is to glean what is learned from existing ML models (trained for different learning tasks and/or DBs and/or workloads), 
and adapt them for new tasks and/or DBs, and/or workloads, while avoiding the need to retrain models from scratch.

{\bf OOD Data insertions.} In analytical DBs data updates primarily take the form of new data insertions. New data may be OOD (representing new knowledge --  distributional shifts), rendering previously-built ML models obsolete/inaccurate.
Or, new data may not be OOD. In the former case, the model must be updated and it must be decided how the new data could be efficiently reflected in the model to continue ensuring accuracy.
In the latter case, it is desirable to avoid updating the model, as that would waste time/resources.
Therefore, it is also crucial to check (efficiently) whether the new data render the previously built model inaccurate. 
However, related research has not yet tackled this problem setting, whereby 
\textit{models for the same learning tasks (e.g., AQP, DG, CE, etc.) trained on old data, continue to provide high accuracy for the new data state} (on old and new data, as queries now may access both old data and new data, old data, or simply the new data).
Related work for learned DB systems have a limited (or sometimes completely lack the) capability of handling such data insertions (as is independently verified in \cite{wang2020we} and will be shown in this paper as well).

{\bf Sources of Difficulty and Baselines.} 
In the presence of OOD, a simple solution is adopted by some of the learned DB components  like Naru \cite{yang2019deep}, NeuroCard \cite{yang2020neurocard}, DBest++ \cite{ma2021learned}, and even the aforementioned transfer/few-shot learning methods \cite{wu2021unified, hilprecht2021one}. That is to "fine-tune" the original model $M$ on the new data. Alas, this is problematic. For instance, while a DBest++ model on the "Forest" dataset has a 95th percentile q-error of 2, updating it with an OOD sample using fine-tuning increases the 95th q-error to ~63. A similar accuracy drop occurs for other key models as well -- \cite{wang2020we} showcases this for learned CE works.
This drastic drop of accuracy is due to the fundamental problem of \catforget{} \cite{mccloskey1989catastrophic}, where retraining a previously learned model on new tasks, i.e. new data, causes the model to lose the knowledge it had acquired about old data. To avoid \catforget{}, Naru and DBest++ suggest using a smaller learning rate while fine-tuning with the new data. This, however, causes another fundamental problem, namely \textit{intransigence}, \cite{chaudhry2018riemannian} whereby the model resists fitting to new data, rendering queries on new data inaccurate.

Another simple solution to avoid these problems would be to aggregate the old data and new data and retrain the model from scratch. However, as mentioned, this is undesirable in our environment. As a concrete example, training Naru/NeuroCard on the "Forest" dataset (with only 600k rows) on a 40-core CPU takes ca. 1.5 hours. Similarly high retraining overheads are typically observed for neural network models, for various tasks.
And, retraining time progressively increases as the DB size increases. 

Therefore, more sophisticated approaches are needed, which can avoid \textit{intransigence} and \catforget{},
update models only when needed and do so while ensuring much smaller training overheads than retraining from scratch and at the same time ensure high accuracy for queries on old and new data. While for some tasks, like CE, some researchers question whether achieving very high accuracy through learned models will actually help the end-task (query optimization) \cite{marcus2021bao}, for tasks like AQP (which is itself the end-task) and for DG (with classification as the end-task) high accuracy is clearly needed, as shown here. Even for CE, with OOD data, accuracy can become horribly poor, as shown here, which is likely to affect query optimization.

\subsection{Contributions} \label{contribution}
To the best of our knowledge, this work proposes the first updatability framework (DDUp) for learned DBs (in the face of new data insertions possibly carrying OOD data)
that can ensure high accuracy for queries on new and/or old data. 
DDUp is also efficient and 
it can enjoy wide applicability, capable of being utilized for different NNs and/or  different learning tasks (such as AQP, DG, CE, etc.). DDUp consists of a novel OOD detection and a novel model-update module. More specifically, the contributions of DDUp are:

\begin{itemize}[leftmargin=10pt]
    \item A general and principled two-sample test for OOD detection. Generality stems from it being based on the training loss function of the NNs. Compared to prior art, it introduces no extra costs and overheads, and could be used with different NNs, with different loss functions, in different applications. To further minimize detection time, it is divided into offline and online phases.
    \item A novel and general formulation of transfer-learning based on sequential self-distillation for model updating. This formulation allows a higher degree of freedom in balancing tasks w.r.t new and old data, can adapt to different models and tasks, and maximizes performance via self-distillation.
    \item Importantly, DDUp can be used by any pre-trained NN without introducing any assumptions on models or requiring additional components that might require to retrain models or incur more costs. Here, we instantiate it for three different tasks (namely, the CE task, using the Naru/NeuroCard deep autoregressive network (DARN) models  \cite{yang2019deep, yang2020neurocard},  the AQP task, using the DBEst++ mixture density network (MDN) model \cite{ma2021learned}, and for the DG task, using the Tabular Variational AutoEncoder (TVAE) model \cite{xu2019modeling}) each of which employs a different NN type. These are representative learning tasks and networks with evident importance in DBs and beyond. These instantiations are also novel, showing how to distil-and-update MDNs, DARNs, and TVAEs.
    \item Finally, DDUp is evaluated using six different datasets and the three instantiated learned DB components, for AQP, CE, and DG
\end{itemize}

\subsection{Limitations} \label{limits}
DDUp focuses only on data insertions, which are essential and dominant in analytical DBs, and not on updates in place and deletes, which are prevalent 
in transactional DBs.
Nonetheless, the latter touch upon an open problem in the ML literature, namely $"unlearning"$, 
where it typically concerns privacy (e.g., removing sensitive data from images in classification tasks) 
(e.g., \citep{sekhari2021remember, golatkar2020eternal}). 
Studying unlearning for DB problem settings is a formidable task of its own and of high interest for future research.

Also, DDUp is designed for NN-based learned DB components. This is so as neural networks are a very rich family of models which have collectively received very large attention for learned DBs. Extending DDUp principles beyond NN models is also left for future research.

\section{The Problem and Solution Overview} \label{problemdef}
\subsection{Problem Formulation} \label{problemformulation}
Consider a database relation \(R\) with attributes \(\{A_1, A_2, ..., A_m\}\). This can be a raw table or the result of a join query. Also consider a sequence of \(N\) insertion updates denoted by \(I=\{I_1,I_2,...I_N\}\). Each \(I_t\) is an insert operation which appends a data batch \(D_t=\{(A_1, A_2, ..., A_m)_t^{(i)}; i=1,..., n_t\}\) to \(R\), where \(n_t\) is the number of rows. Let \(S_t\) be a sufficient sample of \(D_t\) and \(S^{\leq}_{t-1}\) be a sufficient sample from \(\cup_{j=0}^{t-1} D_j\). We naturally assume that \(|R|\) is finite. 
And, due to the training restrictions of existing models, we also make the natural assumption:
\[\forall A_i \in R: supp(D_{t}(A_i)) \subseteq supp(D_{t-1}(A_i)) \]
where \(supp(D(A_i))\) is the support of attribute \(A_i\) in dataset \(D\). This assumption satisfies the condition based on which the domain of each attribute is not violated in the upcoming update batches. 

\textbf{Statistical test for data changes}. We define out-of-distribution detection as a two-sample hypothesis test between a sample of historical data and a sample of the new data. Let \(S^{\leq}_{t-1}\) have a joint distribution of \(P(A_1,\dots, A1_m) \equiv \mathbb{P}\) and \(S_{t}\) have a joint distribution of \(Q(A_1,\dots, A_m) \equiv \mathbb{Q}\). We define the null hypothesis \(H_0: \mathbb{P}=\mathbb{Q}\) which asserts that \(S_{t}\) and \(S^{\leq}_{t-1}\) are coming from a same distribution; and the alternative hypothesis \(H_A: \mathbb{P}\neq \mathbb{Q}\) which declares that the two samples are generated by two different distributions. 

\textbf{Incrementally updating the model}. Consider for \(I_0\) a model \(M_{0}\) is trained by minimizing a loss function \(\mathscr{L}(D_{0};\Theta_0\)). This model may be stale for \(I_t; t>0\). Ideally, the goal of incremental learning is: at time \(t\) train a model \(M_{t}\) that minimizes a function over \(\sum_{i=1}^{t} \mathscr{L}(D_{i};\Theta_i)\). This new model should not forget  \(\{I_{i}; i=0,1,...,t-1\}\) and also learn \(I_t\).

\subsection{A High Level View of DDUp}\label{highlevel} 
\begin{figure*}
  \centering
  \includegraphics[width=0.9\linewidth, height=4cm]{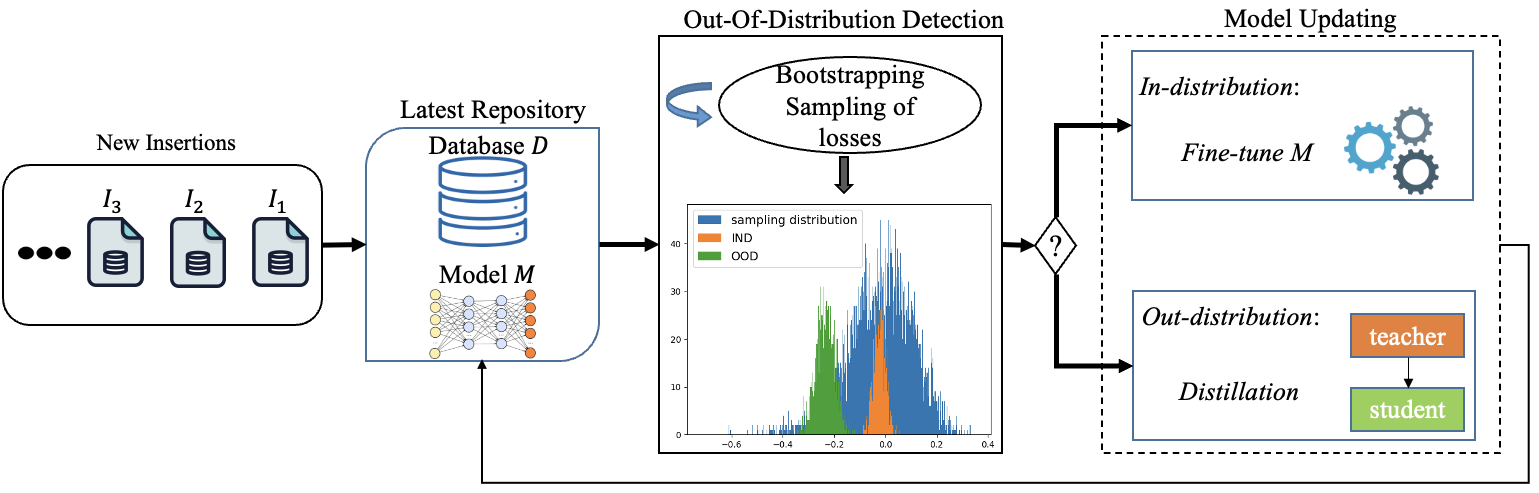}
  \caption{The overall structure of DDUp. DDUp uses the latest model and previous data to build a sampling distribution for the two-sample test, and updates the learned component based on the shift in the data distribution.}
  \label{fig:Arch}
  \vspace{-0.3cm}
\end{figure*}

The overall architecture of DDUp is depicted in \autoref{fig:Arch}.

DDUp process batches of tuples at a time. Such batched handling of insertions is typical in analytical DBs. Furthermore, this takes into account that 
the effect of single tuples is usually negligible for the overall large space modelled by NNs. And, for most tasks like CE, AQP and DG, the effect of single tuples in the final result is very small, considering the large sizes of tables. And batching amortizes detect-and-update costs over many insertion operations.

Upon a new batch insertion, DDUp takes the latest model \(M_{t-1}\), and performs a bootstrapping sampling from the previous data to build the sampling distribution for the average loss values. DDUp uses this distribution to calculate a significance level corresponding to a confidence interval (e.g a 95th confidence interval). The general idea is that if the new data is similar to the previous data (IND in \autoref{fig:Arch}), the loss values of \(M_{t-1}\) for this new data should lie within the threshold. This means that the new data has the same distribution and therefore the model could be left intact (updating maybe just the hyper-parameters of the system, including possible frequency tables and other table statistics. Alternatively, a simple fine-tuning can be performed to adapt the model to the new data.

If the loss values exceeded the threshold, this implies that the data distribution has significantly changed. DDUp will deploy a teacher-student transfer learning method based on knowledge distillation to learn this new distribution without forgetting the knowledge of the old data. In this framework, while the student directly learns the distribution of the new data, the teacher act as a regularizer to make the student also learn about the old distribution.
\vspace{-0.3cm}
\section{Out-of-Distribution Detection} \label{driftdetect}
\subsection{Background} \label{oodback}
In ML, OOD is typically addressed from a classification perspective. Formally, assume \(D\) is a dataset of \((x,y)\) pairs which are drawn from a joint distribution, \(p(x,y)\), where \(x \in \mathcal{X} := \{x_1, x_2, \dots, x_n\}\) is the input (independent variable) consisting of \(n\) features, and \(y \in \mathcal{Y} := \{1,2, \dots, k\}\) is the label corresponding to one of the \(k\) in-distribution classes. A sample \((x,y)\), that probably is generated by a different distribution than \(p(x,y)\), is called OOD, if \(y \notin \mathcal{Y}\), i.e it does not belong to any previously seen classes. 
 
A similar problem has previously been addressed in statistics as {\it concept drift} detection, where different types of shifts are distinguished by expanding \(p(x,y)\) using the Bayes rule:
\begin{equation}\label{bayesrule}
 p(x,y)=p(x)p(y|x)
\end{equation}
Based on Eq. \ref{bayesrule}, changes in \(P(y|x)\) are usually referred to as \textit{Real drift}, while changes in \(P(x)\) are called \textit{virtual drift} \cite{gama2014survey}. In \(X\rightarrow y\) problems the latter mostly is known as \textit{covariate shift}.
Deciding which drift to detect is dependent on the underlying models. For example, deep autoregressive networks (e.g., used by \cite{yang2019deep}) learn the full joint distribution of a table. Hence, they are sensitive to \textit{covariate shift} upon insertions. 
On the other hand, mixture density networks (e.g., used by \cite{ma2021learned}), model the conditional probability between a set of independent attributes and a target attribute. Hence, for these models, one would be interested in detecting \textit{real shift}.
\vspace{-0.2cm}
\subsection{Loss based OOD Detection} \label{llfordrift}
There are several challenges that make it difficult to simply adopt one of the OOD detection algorithms in the ML or statistical learning literature.
First, DB tables are multivariate in nature and learned models are usually trained on multiple attributes. As a result, uni-variate two-sample tests like Kolmogorov–Smirnov (KS) test are not suitable for this purpose. Second, the test should introduce low overheads to the system as insertions may be frequent. Therefore,  multivariate tests like kernel methods that require to learn densities and perform expensive inference computations are not desirable. Third, we aim to support different learning tasks for which different models might be used. Thus, most of OOD detection methods in ML that are based on probability scores (confidence) of classification tasks are not useful here. Moreover, the test should be able to adapt efficiently to the case where insertions occur within old data, that is, without having to recalculate baseline thresholds etc.

An efficient OOD detection method is now proposed that resolves all above issues by leveraging the underlying ML models themselves. Central to most learned data system components is the ability to derive from the underlying data tables a model for the joint or conditional data distribution like \(p(x)\) or \(p(y|x)\). A model usually achieves this by learning a set of parameters \(\Theta\) that represent a function \(f\) by iteratively optimizing over a loss function as follows:

\begin{equation} \label{generalopt}
    f_\Theta = \argmin_{f \in \mathcal{F}} \frac{1}{n} \sum_{i=1}^n \mathscr{L}(f(x);\Theta) + \Omega(f)
\end{equation}

where, \(\Omega\) is a regularizer term, \(n\) is the number of samples, and \(f\) could be the outputs of the model in the last layer (called \textit{logits}), or the probabilities assigned by a "softmax" function.

We will later discuss different loss functions in more details when instantiating different models. In general, loss functions are usually highly non-convex with many local mimina. However, a good learning strategy will find the global minimum. Because of the large data sizes, training is usually done by iterating over mini-batches and a gradient descent algorithm updates the parameters based on the average of loss of the samples in each mini-batch per iteration. For the rest of the paper, when we mention 'loss value' we mean average of losses of the samples in a batch. Once the model is trained, i.e. the loss values have converged, the model can serve as a transformer to map (high-dimensional) input data to the one-dimensional loss functions space around the global minimum. Accordingly, the previous data (seen by the model) are closer to the global minimum compared to the out of distribution data.

The above discussion explains the possibility to compare in- and out-of distribution data just by relying on the underlying models without any further assumptions/components, in a low-dimensional space. With these in hand, we can perform a statistical testing to compare the loss values of old data and new data. In the following we will explain a two-sample test for this purpose. 

\subsection{A Two-Sample Test Procedure}
The steps for a two-sample hypothesis test are: 
1. Define the null, \(H_0\), and alternative hypothesis, \(H_A\). 
2. Define a test statistic \(d\) that tests whether an observed value is extreme under \(H_0\). 
3. Determine a significance level \(\delta\in[0,1]\) that defines the \(type\mhyphen1\ error\) (false positives) of the test. 
4. Calculate \(p\mhyphen value\) which equals the probability that a statistical measure, e.g. distance between two distributions, will be greater than or equal to the probability of observed results.
5. If \(p\mhyphen value <= \delta\) then the \(p\mhyphen value\) is statistically significant and shows strong evidence to reject \(H_0\) in favor of \(H_A\). Otherwise, the test failed to reject \(H_0\). 

The main challenge herein is how to calculate the test significance of the test statistic, i.e the \(p\mhyphen value\). As explained in Section \ref{problemdef}, we aim to detect if the new data that is inserted to the system at time \(t\) has a different distribution than the previous data. Consider \(S_{t-1}^{\leq}\) be a sample of the previous data and \(S_{t}\) be a sample of the newly inserted data. Let \(d(S_{t-1}^{\leq},\ S_{t})\) be a distance function that measures the distance between the two samples. 
Then, the test significance would be \(p\mhyphen value=P(d < d_t | H_0)\) where $d_t$ is our test threshold.

\textbf{Choosing the test statistic}. The test statistic should reflect the similarity of new data to old data. According to our discussion in Section \ref{llfordrift}, we use
the loss function values after the convergence of the models. We use a linear difference between the loss values of the two samples as our test statistics as follows:
\begin{equation}\label{teststatistic}
d(S_{t-1}^{\leq},S_{t}) = \frac{1}{|S_{t-1}|}\sum_{s\in S_{t-1}}\mathscr{L}(s;\Theta) - \frac{1}{|S_t|}\sum_{s\in S_{t}}\mathscr{L}(s;\Theta)
\end{equation}

where \(\mathscr{L}\) is a loss function achieved by training model \(M\) with parameters \(\Theta\). From Eq. \ref{teststatistic} follows that if the loss function is Negative Log Likelihood, and the likelihoods are exact, the test statistic will be the logarithm of the well-known \textit{likelihood-ratio} test. Eq. \ref{teststatistic} also gives intuition about the effect size:  the larger \(d\) is, the larger the difference between two data distributions would be.
Although many of the learned DB models are trained by maximizing likelihood, some other models (e.g., regressions) are trained using a \textit{Mean-Squared-Error} objective. It has been shown \cite{watkins1992maximum} that MSE optimization maximizes likelihood at the same time. Therefore, the form of the distance function in Eq. \ref{teststatistic} still holds. 
The important consequence of Eq. \ref{teststatistic} is that, under i.i.d assumptions for both samples, it can be shown that the central limit theorem holds for distribution of \(d\) under the null hypothesis,
hence, it has a normal limiting distribution with a mean at 0 and unknown standard deviation. 
To estimate the standard deviation (std), we utilize a bootstrapping approach.

\subsection{Offline and Online Steps}

The main performance bottleneck of such an OOD detection is bootstrapping. Fortunately, this part could be performed offline before data insertion. In the offline phase, 
we draw \(n\) bootstrap samples of size \(|S^{\leq}_{t-1}|\) from \(S^{\leq}_{t-1}\). (In practice, when we have access to the original data, we make $n$ bootstrap samples of size $|S_{t-1}^{\leq}|$ from $D_{t-1}^{\leq}$). We use the model \(M_{t-1}\) to compute the average likelihoods (or other losses) of each sample and create a sampling distribution of said average likelihoods. Then, we calculate the standard deviation of the sampling distribution, \(std\), which we will use to find the significance level.


In the online phase, when an insertion happens, we take a sample of the new data, 
\(S_{t}\) and use the latest model, \(M_{t-1}\), to calculate the average likelihood of \(S_{t}\). Finally, we compare the test statistic $d$ with the threshold ($ 2 \ \times \ std$). 
Given the normality of the above distribution of average likelihoods, we know that if one were to draw another sample \(S_{t-1}^{\leq}\) from the previous data its average likelihood would fall within $ 2 \ \times \ std$ from the mean with probability 95\%. Now, if \(d > 2\times std\) we conclude that we are not confident enough to accept that the new data has the same distribution of the old data -- that is, we reject the null hypothesis with a p-value of 0.05.


\subsection{The Test Errors}\label{testerrors}
There are two errors associated with a hypothesis testing. \textit{type-1 error} is rejecting the null hypothesis when it should not.  \textit{Type-2 error} is the error of accepting the null hypothesis when it should be rejected. The first one introduces false positives to the system and the second causes false negatives. 
False positives (FPs) are only a (rather small) performance concern only, spending time to update the model while accuracy is preserved. False negatives (FNs), however, can cause a loss of accuracy.  
Therefore, the system can afford to be stricter with respect to the significance level, in order to reduce the risk of false negatives and accuracy loss.

DDUp uses the loss of the trained NNs for OOD detection. Sometimes NNs could be over-confident \cite{nguyen2015deep,ren2019likelihood,nalisnick2018deep} which may introduce bias. 
However, we have not witnessed it for our tasks here on tabular data.
If there were bias, the FP and FN rates discussed above would signal it. 
We have evaluated DDUp with respect to FPs/FNs in Section \ref{oodeval} showing that this is not a concern.

\section{Model Update} \label{KD}
In this section, we propose a transfer-learning based method that can retain previous knowledge of the model while adapt it to the new insertions. The OOD detection module will either output 'in-distribution' or 'out-of-distribution' signals.

\textbf{The in-distribution case}. When no drift occurs, the new data distribution is similar to that of the historical data and this distribution could be represented by a similar parameter space of the latest model, \(M_{t}\). 
Hence, the learned component of the system could remain unchanged. More specifically, the framework can copy \(M_{t}\) to \(M_{t+1}\) and update the required meta-data associated with the system (such as the frequency tables in DBEst++, or table cardinalities in Naru/NeuroCard).  Even if there are slight permutations in data, fine-tuning the latest model's parameters on the new data will adjust it to the general representation of both old and new data. 
We will show that when knowing that data is not OOD, \finetuning{} with a relatively small learning rate, can retain model performance. 
Specifically, with an \textbf{in-distribution} signal at time \(t+1\), \(M_{t}\) is retrained on \(S_{t+1}\) with a small learning rate, $lr$. This learning rate could be tuned, as a hyper-parameter. 
We intuitively set \(lr_{t} = \frac{|D_{t+1}|}{|D_{t}^\leq|}\ \times \ lr_{0}\) and experimentally show that it is a good choice. 

\textbf{The OOD case}. With a distributional shift,
by fine-tuning on new data, the model's parameters would bias toward the new data distribution. Even smaller learning rates cause tiny deviations from the previous parameter space which may yield large errors during inference. And, retraining using all the data from scratch is too time consuming. Thus, we propose an updating approach grounded on the transfer-learning paradigm. The general idea is to use the learned model \(M_{t}\) and incorporate it in training \(M_{t+1}\). To this end, we utilize the \kndistill{} principles, which help to transfer the previously learned knowledge to a new model. Our rationale for such a model updating approach is based on the following: 
\begin{itemize}[leftmargin=*]
    \item Distillation has several benefits including: faster optimization, better generalization, and may even outperform the directly trained models. \cite{yim2017gift}.
    \item It is accurate for queries on old  as well as new data.
    \item It allows us to control the weights for queries on new and old data with just a couple of parameters.
    \item It is efficient memory-wise as well as computationally-wise, compared to methods like Gradient Episodic Memory, or Elastic Weight Consolidation and PathInt (cf. Section \ref{litraturere})
    \item It does not make any assumptions about the training of the underlying models. This property, is especially desirable since: a) we can use it to update different neural networks; b) it prevents the high costs of rebuilding base models; c) different pre-processings could be left intact. For instance, Naru, DBEst++ and TVAE all use completely different types of embedding/encoding. DDUp can update the model regardless of these differences.
\end{itemize}

\subsection{General Knowledge Distillation (KD)}
\begin{figure}[hb]
    \centering
    \includegraphics[width=\linewidth]{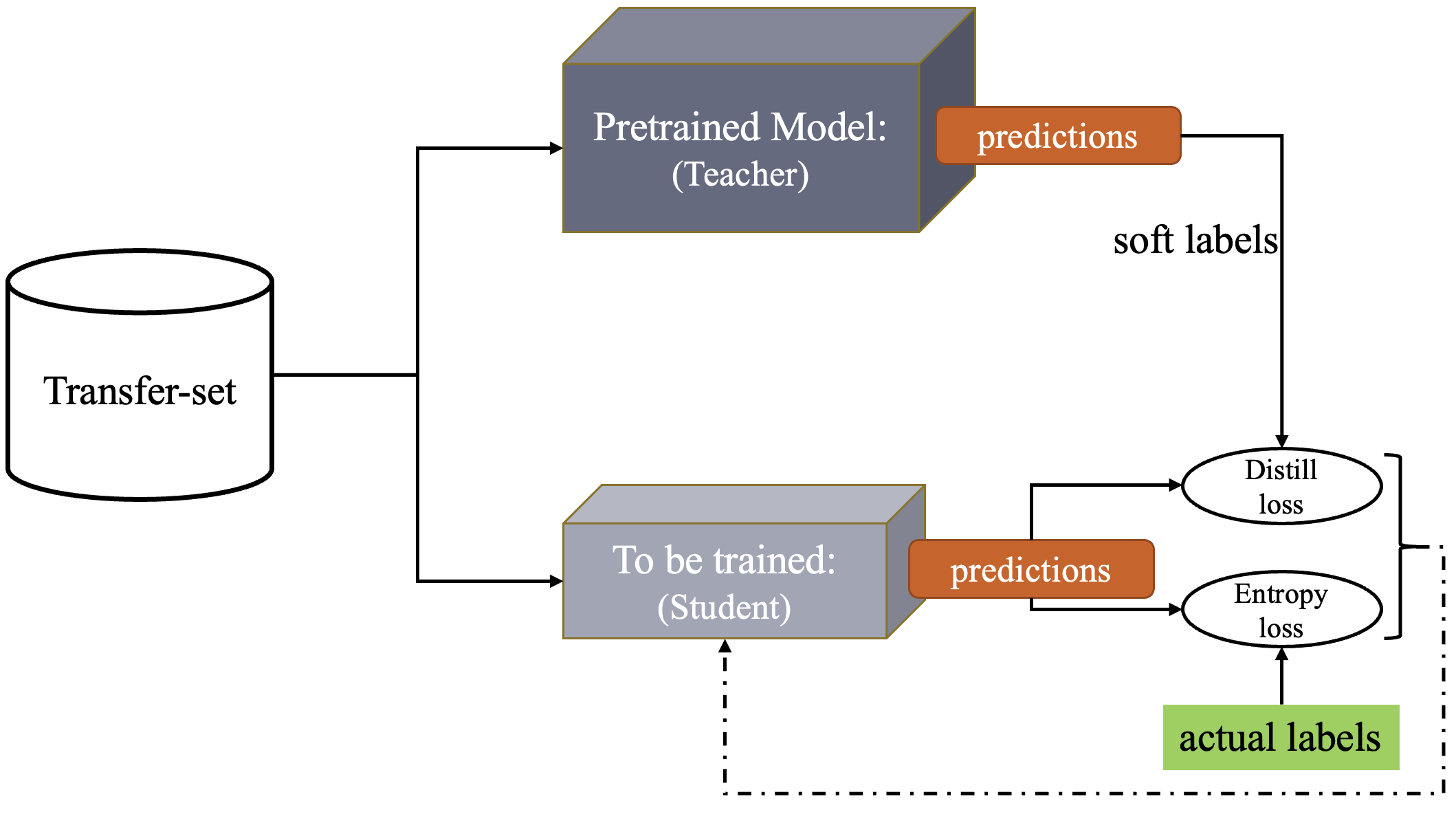}
    \vspace{-0.2cm}
    \caption{The knowledge distillation process.}
    \label{fig:kdfig}
    \vspace{-0.35cm}
\end{figure}

KD was first introduced in \cite{hinton2015distilling} for $model \  compression$ by transferring knowledge from an accurate and "cumbersome" model, called \textit{teacher}, to a smaller model called \textit{student}. In its basic form, instead of fitting the student model directly to the actual data \textit{labels}, one would use the class probability distribution learned by the teacher to fit the student model. Hinton et al. \cite{hinton2015distilling} argued that small probabilities in "wrong" label logits, known as "soft labels", include extra information called "dark knowledge" that result in better learning than actual "hard labels". Distillation has since been extensively studied. \autoref{fig:kdfig} shows a general view of the principles of a distillation process. A small dataset referred to as \textit{transfer-set} is fed into a pre-trained model (teacher) and a new model (student) to be trained. A $distillation \ loss$ is calculated using the predictions of the pre-trained model instead of the actual labels. This loss and a typical loss using actual labels will be used to train the new model. 

To formulate \kndistill{}, consider a model with parameters \(\Theta\), representing a function \(f_t\) (\(t\) for teacher) which has been trained via Eq. \ref{generalopt}. We would like to transfer knowledge from this teacher model to a student model with parameter \(\Theta'\), representing a function \(f_s\). This new model could be trained as follows:

\begin{equation} \label{distillopt}
    f_{s\Theta'} = \argmin_{f \in \mathcal{F}} \frac{1}{|tr|} \sum_{i\in tr} \left[\lambda\mathscr{L}_d(f_s(i);f_t(i);\Theta;\Theta') + (1-\lambda)\mathscr{L}(f_s(i);\Theta')\right]
\end{equation}
\\
for weight \(\lambda\), distillation loss \(\mathscr{L}_d\), and transfer-set \(tr\). 

\subsection{DDUp: Updating By Knowledge Distillation}\label{upbykd}

\cite{furlanello2018born,seq-self-distill} showed that, for classification tasks, if instead of having a compact student model, one uses the same architecture of the teacher, and repeat distillation sequentially for several generations, the student models in the later generations could outperform the teacher model. This approach is called {\it sequential self-distillation}.
Inspired by this and anticipating that this will be valid for our learning tasks, DDUp also employs a sequential self-distillation approach.

To update a model using KD, 
a copy of the previously trained model becomes the new student. Then,  the student is updated using a distillation loss (to be defined soon). After updating, the previous teacher is replaced with the new updated model. This cycle repeats with every new insertion batch. 

To formulate our training loss function, we consider two aspects that we would like to have in our updating scheme. First, to have control over the the new data/queries versus the old data/queries. Second, to make it general so that different learned DB systems could adopt it. As such, we first write down the general form of the total loss function and then, use cross-entropy and mean-squared-error as the loss functions to instantiate different models. Training in each update step is as follows:
\\
\begin{equation} \label{totalloss}
\begin{split}
    f_{s\Theta'} = \argmin_{f \in \mathcal{F}} & \bigg(\alpha \times \frac{1}{|tr|} \sum_{x\in tr} \big[\lambda\mathscr{L}_d(f_s(x),f_t(x);\Theta') \\ 
    & + (1-\lambda)\mathscr{L}(f_s(x);\Theta')\big] \\
    & + (1-\alpha) \times \frac{1}{|up|}\sum_{x\in up}\mathscr{L}(f_s(x);\Theta') \bigg)
\end{split}
\end{equation}
\\
Here, \(\alpha\) and \(\lambda\) are the new data and the distillation weights, respectively. Also, \(tr\) and \(up\) are the transfer-set and the update batch. 
In summary, the rationale for proposing this novel loss function is: 
The transfer-set term acts as a regularizer to avoid overfitting on new data.
The same goal is also helped by self-distillation (when copying the teacher to the student). Additionally, as mentioned sequential self-distillation \cite{seq-self-distill} may attain increasingly higher accuracy, even outperforming "retrain from scratch"
(cf. Section 5.3). 

For models that provide a conditional probability in the last layer of the network (e.g. using a Softmax function), an annealed cross-entropy loss will be employed. Otherwise, we utilize mean-squared-error using the logits from the last layer of the network. Eq. \ref{cedistillloss} and Eq. \ref{mseloss} show these two loss functions. 
\\
\begin{equation}\label{cedistillloss}
    \mathscr{L}_{ce}(D_{tr};z_t,z_s) = - \sum_{i\in [k]} \frac{exp(z_{t_i}/T)}{\sum_{j\in [k]}exp(z_{t_j}/T)} \log \frac{exp(z_{s_i}/T)}{\sum_{j\in [k]}exp(z_{s_j}/T)}
\end{equation}
\\
\begin{equation} \label{mseloss}
    \mathscr{L}_{mse}(D_{tr};z_t,z_s) = \sum_{i\in[|z_t|]}(z_{t_i} - z_{s_i})^2
\end{equation}
\\
where \(D_{tr}\) is the \textit{transfer-set} , \(T\) is a temperature scalar to smooth the probabilities so that it produces "softer" targets, and \([k]\) is the vector \([0,1,\dots,n]\) which are the class probabilities and \([|z_t|]\) indicates the logits of the network.

\subsection{Instantiating the Approach}

\textbf{Mixture Density Networks}. MDNs consist of an NN to learn feature vectors and a mixture model to learn the \textit{probability density function} (pdf) of data. Ma et al. \cite{ma2021learned} uses MDNs with Gaussian nodes to perform AQP. For the Gaussian Mixture, the last layer of MDN consists of three sets of nodes \(\{\omega_i, \mu_i, \sigma_i\}_{i=1}^m\) that form the pdf according to Eq. \ref{mdneq}. 

\begin{equation}\label{mdneq}
\hat{P}(y|x_1, ..., x_n) = \sum_{i=1}^{m}\omega_i.\mathscr{N}(\mu_i, \sigma_i)    
\end{equation}

where \(m\) is the number of Gaussian components, \(y\) is the dependent variable and \((x_1, ..., x_n)\) is a set of independent variables, \(w_i\) is the weight of the \(i^{th}\) Gaussian with a mean of \(\mu_i\) and a standard deviation of \(\sigma_i\). 
For MDNs, we define distillation loss as follows:

\begin{equation} \label{mdnkdloss}
\mathscr{L}_d = \mathscr{L}_{ce}(D_{tr}, \omega_{t}, \omega_{s}) + \mathscr{L}_{mse}(D_{tr}, \mu_{t}, \mu_{s}) + \mathscr{L}_{mse}(D_{tr}, \sigma_{t}, \sigma_{s})
\end{equation}

This summation of terms help us retain both the shape of data distribution as well as the intensity levels. 

\textbf{Deep Autoregressive Networks}. The Naru and NeuroCard cardinality estimators \cite{yang2019deep, yang2020neurocard} use deep autoregressive networks (DARNs) to approximate a fully factorized data density. DARNs are generative models capable of learning full conditional probabilities of a sequence using a masked autoencoder via Maximum Likelihood. Once the conditionals are available, the joint data distribution could be represented by the product rule as follows:
\[
\hat{P}(A_1, A_2, \dots, A_n) = \hat{P}(A_1)\hat{P}(A_2|A_1)\dots \hat{P}(A_n|A1,\dots ,A_{n-1})
\]

where \(A_i\) is an attribute in a relation \(R\). Naru and NeuroCard use cross-entropy between input and conditionals as the loss function. This allows us to formulate the distillation loss function using the conditionals of the teacher and the student networks. Also, in Naru and NeuroCard, each conditional is calculated using a set of logits, hence we average over all as follows:
\begin{equation} \label{narukdloss}
\mathscr{L}_d = \frac{1}{|A|}\sum_{i=1}^{|A|}\mathscr{L}_{ce}(D_{tr}, z_{s_i}, z_{t_i})
\end{equation}

Where \(|A|\) is the number of attributes corresponding to the number of conditionals.

\textbf{Variational Autoencoders}. VAEs have been used for a number of DB components: \cite{thirumuruganathan2020approximate} for AQP, \cite{hasan2020deep} for CE, and \cite{xu2019modeling} for synthetic tabular data generation. 
They are a type of autoencoders that instead of learning deterministic encoder, decoder, and compressed vector (known as bottleneck), they learn a probabilistic encoder, decoder, and a latent random variable instead of the compressed vectors. (For more details, see the seminal paper \cite{kingma2013auto}). 
Interestingly, a VAE is trained using a different loss function, known as Evidence-Lower-Bound (ELBO) loss (which amounts to a lower bound estimation of the likelihoods).
Here we shall use TVAE for learned synthetic tabular data generation (of particular importance in privacy-sensitive environments, or when data is scarce for data augmentation purposes, or when wishing to train models over tables and accessing raw data is expensive in terms of time or money).

To distill a VAE, one must cope with the random noise added to the input of the decoder by the latent variable. For that, the latent variable in the teacher network is removed, and we use the same noise generated by the student in the teacher. The reason for doing this is that distillation tries to teach the student to behave like the teacher for a specific observation or action. If there is randomness, the student might mimic the teacher's behaviour for a completely different observation. After this change, the corresponding logits of the encoder/encoder of the student and the teacher are compared using MSE. Finally, the loss function is:
\\
\begin{equation}
    \mathscr{L}_d = \frac{1}{2}( \mathscr{L}_{mse}(D_{tr}, z_t^{(e)}, z_s^{(e)}) + \mathscr{L}_{mse}(D_{tr}, z_t^{(d)}, z_s^{(d)}) )
\end{equation}
\\
where, \(e\) and \(d\) correspond to the encoder and the decoder networks.

\subsection{An Example}
\begin{figure}
    \centering
    \includegraphics[width=\linewidth]{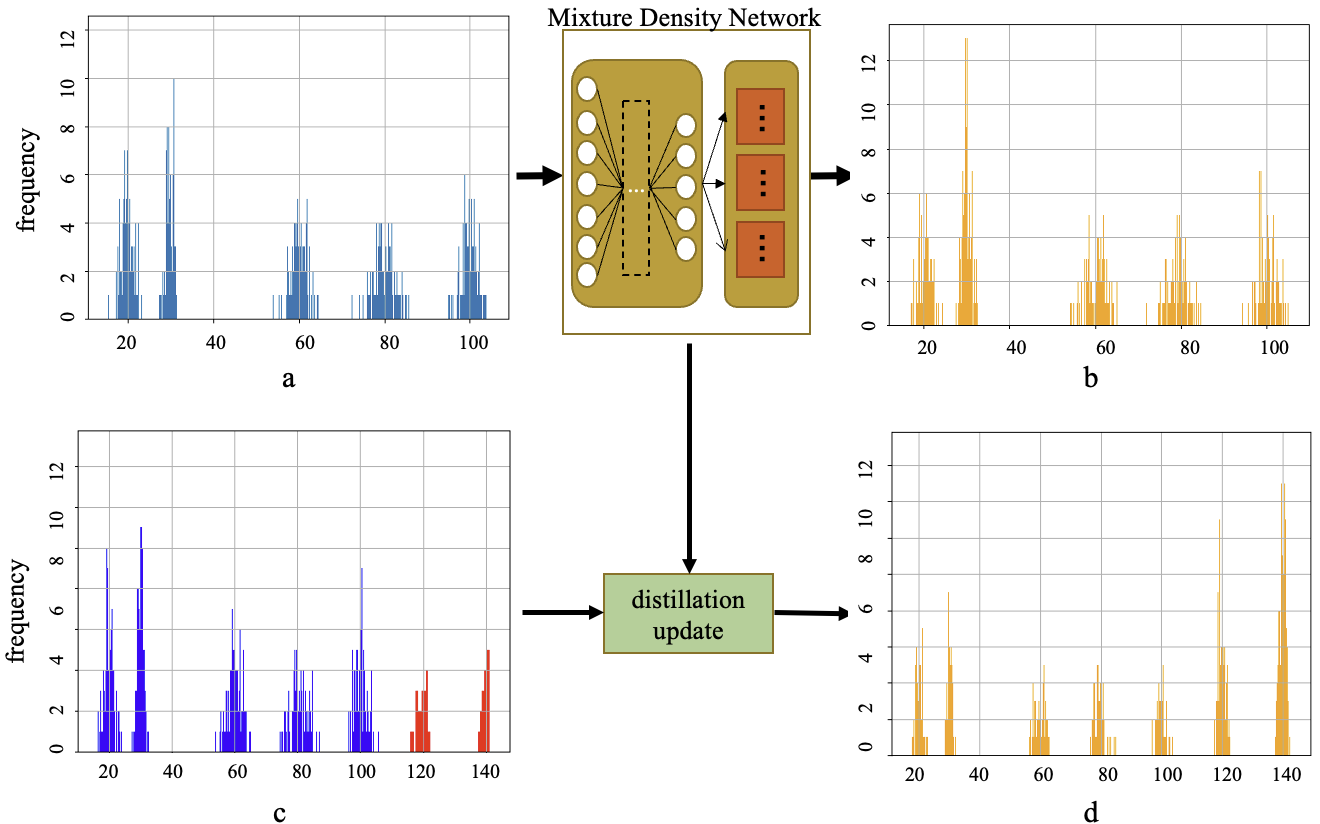}
    \caption{An example to show how DDUp learns new data without forgetting. 'a' is the histogram of synthetic data corresponding to $x=1$. 'b' is the sample generated by the learned MDN for $x=1$. 'c' shows a sample of an update batch coming from different Gaussians. 'd' is the sample generated by the MDN after being updated by the DDUp loss function. We have performed the update 50 times to see the effect of high frequency updates (This explains the higher frequencies around the last two peaks for 'd').}
    \vspace{-0.4cm}
    \label{fig:toyexam}
\end{figure}

We create a simple synthetic dataset consisting of a categorical attribute, \(x\), with 10 \(distinct\mhyphen values=\{1,2,3,\dots,9,10\}\), and with each category having 1000 real values. The dataset is balanced and the real values for each category are generated by a \textit{Mixture of Gaussians} (MoG) with five peaks. \autoref{fig:toyexam}.a is the dataset corresponding to \(x=1\). We fit a \textit{Mixture Density Network} with ten components on this dataset. \autoref{fig:toyexam}.b shows a sample generated by this MDN which asserts that the model has perfectly learnt the data distribution. Next, we introduce an update batch generated by a MoG with two different means. \autoref{fig:toyexam}.c shows the update batches in red color compared to the previous data in blue. We update the previously learned MDN with the proposed loss function in Eq. \ref{mdnkdloss}. We repeat updates for 50 batches generated with the new MoG. \autoref{fig:toyexam}.d shows the final distribution learnt by the MDN. 

\subsection{Handling Join Operations}\label{joins}
DDUp can operate either on raw tables or tables from join results.
If the old data $R$ is the result of a join, the new data batch needs to be computed, due to new tuples being inserted in any of the joined tables in $R$.
Consider at time \(t-1\) a model \(M_{t-1}\) has been trained on  \(R=\bigcup_{j=0}^{t-1}{T_1^j} \bowtie \bigcup_{j=0}^{t-1}{T_2^j} \dots \bowtie \bigcup_{j=0}^{t-1}{T_n^j}\), where $T_r^j$ denotes the new batch for table $T_r$ at time $j$. 
Without loss of generality, suppose a new insertion operation $I_t$ at time \(t\)  adds new data to table \(T_i\), denoted \(T_i^t\). The new data for DDUp in this setting is \( D_t = (R\ \setminus \  \bigcup_{j=0}^{t-1}T_i^{j}) \bowtie T_i^{t} \), where $\setminus$ denotes a (multi)set-difference operator. Therefore, for the detection module, \(S^{\leq}_{t-1}\) is a sample of R, and \(S_t\) a sample from $D_t$. Furthermore, during updating the transfer-set is a sample from $R$ and the new data is $D_t$. 
Please note that all this data preparation and how each model deals with joins is orthogonal to DDUp.
Therefore, it can be done by either computing the actual joins above or using join samplers like \cite{zhao2018random,shanghooshabad2021pgmjoins}, as is done in NeuroCard and compared against in Section \ref{joinexp}.

\section{Experimental Evaluation} \label{eval}
We evaluate DDUp for three different models for learned DB components: (i) Naru/NeuroCard \cite{yang2019deep,yang2020neurocard} which use DARN models for CE; (ii) DBest++ \cite{ma2021learned} that uses MDNs for AQP; and (iii) TVAE \cite{xu2019modeling}, that uses variational autoencoders for DG. 
We evaluate in terms of model accuracy and update time. 
We use as reference points the baseline update approach provided for AQP and CE (TVAE provides no update approach).
We also add as reference points the accuracy when retraining from scratch and when leaving models stale. With respect to \textit{OOD detection}, we investigate whether it can detect significant data shifts successfully and how this will contribute to the final performance of the underlying models in their specific application, CE, AQP, DG. 
Ultimately, the experiments are to address the following questions:

\begin{itemize}[leftmargin=*]
    \item How to best evaluate DDUp? (Section \ref{setup})
    \item Can DDUp accurately detect a distributional shift? (Section \ref{oodeval})
    \item Is DDUp accurate under in- $and$ out-of- distribution settings? (Section \ref{perfeval})
    \item How does DDUp compare to the baseline approaches in accuracy and update time? (Section \ref{perfeval})
    \item What is the effect of distillation? (Section \ref{distilleval})
    \item Is DDUp efficient? (Section \ref{overheads})
\end{itemize}

\vspace{-0.3cm}
\subsection{Experimental Setup} \label{setup}
To establish a dynamic setup, we make a copy of the base table and randomly sample 20\% of its rows as new data. In this setting, new data follows the previous data distribution which we denote as \textit{in-distribution}. We introduce distributional drift as is typically done for tabular data settings, say in \cite{wang2020we}. As such, after making the copy, we sort every column of the copied table individually in-place to permute the joint distribution of attributes. Next, we shuffle the rows and randomly select \(20\%\) of the rows - this now becomes the new data.
With these new data, we perform two types of experiments. First, we consider the whole 20\% sample as a new data batch and update the model with it. Second, to show the updatability in incremental steps, we split the 20\% data into 5 batches. 
In general, the size of the transfer-set is a tunable parameter \cite{hinton2015distilling}, influenced by the dataset complexity, the underlying model generalization ability, and the downstream tasks. 
After tuning, we used a 10\% transfer-set for MDN and DARN and a 5\% for TVAE, which could be further tuned with methods like Grid search.

DDUp does not impose any further constraints to those of the underlying models. For DBest++ we use a query template with a range and an equality attribute. Also, we use one-hot encoding to encode categorical attributes and normalize the range attribute to \([-1,1]\). For Naru/NeuroCard and TVAE, we use the same settings as explained in their code documentation. We use the learned hyper-parameters of the base model, i.e the model we build at time zero, for all subsequent updates. Furthermore, we intuitively set \(\alpha\) parameter in Eq. \ref{totalloss} to the fraction of update batch size to the original data size and tune \(\lambda\) for values in \([9/10, 5/6, 1/4, 1/2]\). 

\subsubsection{Datasets} \label{datasets}
We have mainly used three real-world datasets (census, forest, DMV) 
(see \autoref{tab:Datasets}). These datasets
have been widely used in the learned DB literature. 
For CE, \cite{wang2020we} uses also forest, census and DMV, while NeuroCard/Naru use JOB/DMV. For AQP DBEst++ uses TPCDS.  For DG, \cite{xu2019modeling} uses census and forest. Thus, we have also used census, forest, DMV, and TPCDS (\texttt{store sales} table, scaling factor of 1). Finally, for join queries, we have used JOB (on IMDB data) and TPCH benchmarks, which are also used in \cite{yang2020neurocard, yang2019deep}.

\begin{table}[hb]
  \caption{Characteristics of datasets.}
  \vspace{-0.3cm}
  \label{tab:Datasets}
  \begin{tabular}{c c c c} 
    \toprule
     Dataset&Rows&Columns&Joint Domain\\
    \midrule
     Census & 49K & 13 & $10^{16}$  \\
     Forest & 581K & 10 & $10^{27}$  \\
     DMV & 11.6M & 11 & $10^{15}$  \\
     TPCDS & 1M & 7 & $10^{30}$ \\
    \bottomrule
  \end{tabular}
  \vspace{-0.5cm}
\end{table}

\subsubsection{Workload} \label{workload}Each model is evaluated using 2,000 randomly generated queries. These queries are generated at time zero for each model and are used throughout the subsequent updates. When an update batch is performed, the ground truth of the queries will be updated. For Naru/NeuroCard, we use their generator to synthesize queries: It randomly selects the number of filters per query (forest:[3,8], census: [5,12], TPCDS: [2,6], dmv: [5,12]). Then, it uniformly selects a row of the table and randomly assigns operators \([=,>=,<=]\) to the columns corresponding to the selected filters. Columns with a domain less than 10 are considered categorical and only equality filters are used for them. For DBest++, we select a \(lower\mhyphen bound\) and a \(higher\mhyphen bound\) for the range filter and uniformly select a category from the categorical column for the equality filter. Throughout the experiments, we discard queries with actual zero answer. The structure of a typical query in our experiments is:

\begin{lstlisting}[mathescape=true,
    basicstyle=\footnotesize, %or \small or \footnotesize etc.
]
SELECT AGG(y) FROM $T_1 \bowtie T_2 \dots \bowtie T_n$ WHERE $F_1$ AND ... AND $F_{d}$
\end{lstlisting}

where, \(F_i\) is a filter in one of these forms: \([att_i = val, att_i >= val, att_i <= val]\). Also, \texttt{AGG} is an aggregation function like \texttt{COUNT}, \texttt{SUM}, \texttt{AVG}. For DBest++, the query template contains one categorical attribute and one range attribute. As such, we select the following columns from each dataset: census:[\texttt{age, country}]; forest:[\texttt{slope, elevation}]; dmv:[\texttt{body type, max gross weight}]; TPCDS:[\texttt{ss quantity,ss sales price}]; IMDB:[\texttt{info type id,production year}]; TPCH:[\texttt{order date,total price}] where the first/second attribute is categorical/numeric. Furthermore, Naru could not train on the full TPCDS dataset as the encodings were too large to fit to memory. Hence, we selected the following columns [\texttt{ss sold date sk}, \texttt{ss item sk}, \texttt{ss customer sk},\texttt{ss store sk}, \texttt{ss quantity}, \texttt{ss net profit}], and made a 500k sample.

\subsubsection{Metrics}
For \textit{count} queries, we use \textit{q-error} as follows:

\begin{equation}
    error = \frac{max(pred(q), real(q))}{min(pred(q), real(q))}
\end{equation} 

For \textit{sum} and \textit{avg} aggregates, we use \textit{relative-error} as follows:
\begin{equation}
    error = \frac{|pred(q) - real(q)|}{real(q)}\times100
\end{equation} 

Additionally, Lopez et al. \cite{lopez2017gradient} introduce the notions of Backward Transfer (BWT) and Forward Transfer (FWT) as new metrics in class incremental learning tasks. BWT is the average accuracy of the model on old tasks, and FWT is the average accuracy of the model on new tasks. Here, we re-frame BWT and FWT. 
We generate the queries at time \(0\) and use them for all update steps. At each step \(t\), we calculate \(diff = real_t(q) - real_{t-1}(q)\) for each query, \(q\), which gives us three set of queries; \(G_{fix}\) with \(diff=0\), \(G_{changed}\) with \(diff>0\), and \(G_{all} = G_{fix} \cup G_{changed}\). With these groups, we define three measures. \(AT\): average q-error over \(G_{all}\). \(FWT\): average q-error over \(G_{changed}\). \(BWT\): average q-error over \(G_{fix}\).

\subsubsection{Evaluating Variational Autoencoders}
DG is an  interesting learned application which is recently supported using TVAE. Thus, we evaluate DDUp for TVAE. In TVAE, once the training is done, only the decoder network is kept and used, as this is the generator. Hence, we apply our distillation-update method to the decoder network. We evaluate TVAE via the accuracy of an XGboost classifier trained by the synthetic samples, as in \cite{xu2019modeling}. 
We hold-out 30\% of table as the test set, and train two classifiers with original and synthetic data, then predict the classes of the held-out data. We report \textit{micro f1-score} for classifiers. For census, forest and DMV, we use: \textit{income}, \textit{cover-type}, and \textit{fuel-type}, as the target class, respectively.
For TVAE, we created a smaller DMV with 1m records, as training TVAE on the whole DMV is very time/resource consuming (proving indirectly the need to avoid retraining).

\subsection{OOD Detection} \label{oodeval}

\subsubsection{Loss Functions as Signals}
We first show the results of loss/log-likelihoods when the detector receives samples from the same distributions or from different distributions. The results are shown in \autoref{tab:avgll}. For Naru/NeuroCard and DBEst++ we report the actual log-likelihood values (not negatives, so higher is better). For TVAE, we report the ELBO loss values (hence lower is better). 

\begin{table}[hb]
 \centering
  \caption{Average log-likelihood and ELBO loss values of data samples on a trained model. $S_{old}$ is a sample of the previous training data. "IND", is a 20\% sample from a straight copy of the original table; "OOD", is a 20\% sample from a permuted copy of the original table.}
  \vspace{-0.2cm}
  \label{tab:avgll}
  \resizebox{\linewidth}{!}{%
  \begin{tabular}{c c c c  | c c c | c c c } 
    \toprule
\multirow{2}{*}{Dataset} & \multicolumn{3}{c|}{DBEst++} & \multicolumn{3}{c|}{Naru/NeuroCard} & \multicolumn{3}{c}{TVAE} \\
& $S_{old}$  & IND & OOD & $S_{old}$ & IND & OOD & $S_{old}$ & IND & OOD \\
    \midrule
     Census & -0.362 & -0.361 & -0.366 & -20.99 & -20.87 & -36.95 & -15.21	& -15.22 & 81.47  \\
     Forest & -0.0194 & -0.0202 & -0.052 & -43.16 & -43.9 & -141.10 & -19.96 & -20.09 & 142.38 \\
     DMV & 2.520 & 2.532 & 2.444 & -13.74 & -13.16 & -18.67 & 9.114 & 9.28 & 34.95 \\
    \bottomrule
  \end{tabular}}
  \vspace{-0.35cm}
\end{table}

\autoref{tab:avgll} shows that the loss function (log likelihood and ELBO in our cases) can reliably signal OOD data.
Interestingly, this corroborates similar findings in \cite{detectOOD-iclr17} for classification tasks in various vision and NLP tasks, where the NN outputs can be used to signal OOD. Here we show it for tabular data and for NNs developed for AQP, CE, and DG tasks.  

In Naru/NeuroCard and TVAE, when permuting, all columns are sorted individually, hence the large difference in likelihoods. 
For DBEst++, only the selected columns for a query template have been permuted, yielding a small difference in likelihoods. 

\subsubsection{The two-sample test results} 
\autoref{tab:driftdetect} shows  results for two-sample testing for OOD detection. The significance level of the test (threshold) is \(2\times std\) of the bootstrapping distribution, which was obtained by $>$1000 iterations. 
In each iteration, we use a 1\% sample with replacement from previous data and a 10\% sample without replacement from new data to calculate the test statistic. The results show that when data is permuted, the test statistic is far away from the threshold. This means it appears at a great dissonance in the tails of the bootstrapping distribution. 
And since the critical value to test for OOD is found by bootstrapping over \(S_{old}\), i.e., \(S^{\leq}_{t}\), it will adjust even to small differences when faced with OOD. 
Case in point, the DBEst++ OOD likelihood value for census (which is similar to IND/$S_{old}$ in \autoref{tab:avgll}) vs the corresponding test-statistic value in  \autoref{tab:driftdetect}.

\begin{table*}[t]
  \caption{The test-statistic values. Threshold is  $2\times standard-deviation$ and bs-mean is the mean of bootstrapping distribution. }
  \label{tab:driftdetect}
  \resizebox{\textwidth}{!}{
  \begin{tabular}{c | c c c c  | c c c c | c c c c } 
    \toprule
\multirow{2}{*}{Dataset} & \multicolumn{4}{c|}{DBEst++} & \multicolumn{4}{c}{Naru/NeuroCard} & \multicolumn{4}{c}{TVAE} \\
& bs-mean & threshold & IND  & OOD & bs-mean & threshold & IND & OOD & bs-mean & threshold & IND & OOD \\
    \midrule
Census&-0.3524 & 0.007 & 0.001 & 0.05 & -21.0076 & 0.0529 & 0.032 & 16.0052 & -15.1834 & 0.6041 & 0.0419 & 100.5126 \\
Forest&-0.0228 & 0.0122 & 0.007 & 0.2315 & -41.35 & 0.0141 & 0.0084 & 72.5473 & -19.99 & 0.0868 & 0.0417 & 167.0502 \\
DMV&2.52 & 0.1287 & 0.0145 & 4.5745 & -13.7674 & 0.0012& 0.0007& 5.1145 & 9.1209 & 0.0177 & 0.0015 & 25.1398 \\
    \bottomrule
  \end{tabular}}
\end{table*}


\subsubsection{FP and FN rates in OOD detection}\label{fpfnrates}

To evaluate OOD detection, we measure FP and FN rates (FPR, FNR). 
We created an OOD test-set and an IND test-set, each equaling half the original size of the table. The latter is just a random sample from the original table. The former is constructed as follows. The perturbed data is obtained by perturbing one or more of five columns of the table, say $C1, \ ... \ C5$. First we perturb $C1$ and take a sample of the resulting table of size $10\%$ and append it to the OOD test-set. Then we perturb $C1$ and $C2$ and similarly sample and append it to the OOD test-set. We repeat this for perturbations on $C1, C2, C3$, on $C1, C2, C3, C4$, and on $C1, C2, C3, C4, C5$, ending up with an OOD test-set of size 50\% of the original table. Note that this setup creates a more-challenging case, as the degree of perturbations (for OOD data) is finer-grained.
Then, at each batch, we fed a random sample from the OOD test-set and of the IND test-set to the DDUp detector. For each batch, the detector would signal IND or OOD and we recorded and calculated FPR and FNR. The batch size was 2,000 and we repeated the experiment for 1,000 batches.

We used the same parameters for all datasets and models: the bootstrapping size is 32 and the threshold is \(2 \times std\). For DBEst++, the results are reported in \autoref{tab:fprfnr}.  FPR and FNR for Naru/NeuroCard and TVAE were always zero. These results further confirm that the OOD detection algorithm is not biased.

\begin{table}[hb]
  \vspace{-0.3cm}
 \centering
  \caption{FPR and FNR for DBEst++.}
  \vspace{-0.3cm}
  \label{tab:fprfnr}
  \begin{tabular}{c c c } 
    \toprule
Dataset & FPR & FNR \\
    \midrule
     Census & 0.15 & 0.01 \\
     Forest & 0.10 & 0 \\
     DMV & 0.01 & 0 \\
    \bottomrule
  \end{tabular}
  \vspace{-0.35cm}
\end{table}

\begin{figure}
\begin{minipage}{.5\linewidth}
\subfloat[]{\label{main:a}\includegraphics[scale=.29]{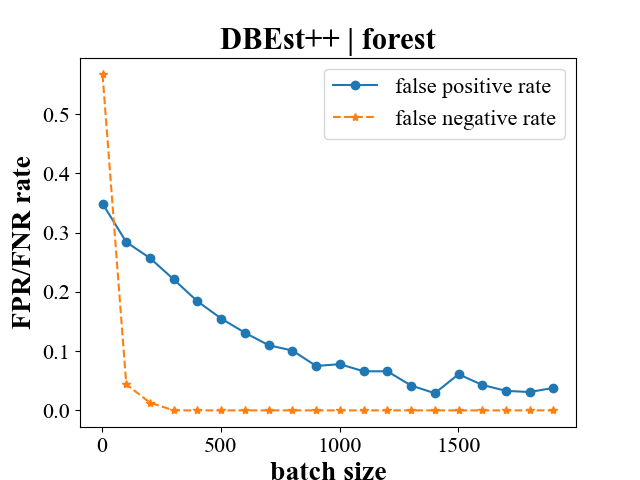}}
\end{minipage}%
\begin{minipage}{.5\linewidth}
\subfloat[]{\label{main:b}\includegraphics[scale=.29]{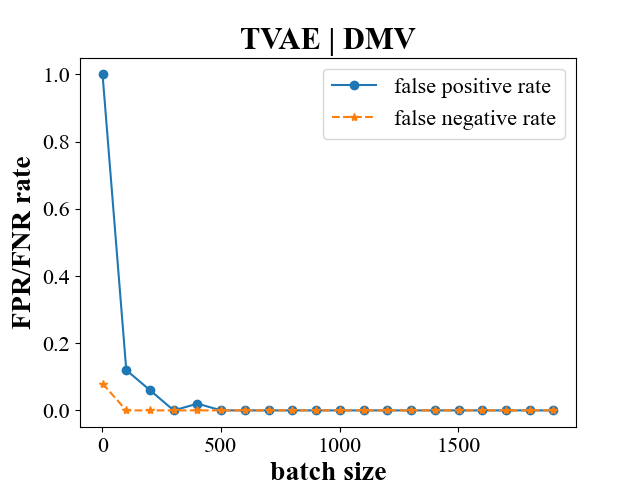}}
\end{minipage} %
\vspace{-0.4cm}
\caption{Sensitivity of OOD detection vs batch size.}
\label{fig:oodsens}
\vspace{-0.55cm}
\end{figure}

Furthermore, we studied the sensitivity on the batch size and varied it from a size of 1 to 2,000. Results are shown in \autoref{fig:oodsens}, which clearly show that after a low-threshold batch size, FPR and FPN tend to zero. The same results hold for other models and datasets, and are omitted here for space reasons.

\subsection{Accuracy Results} \label{perfeval}
\subsubsection{When there is OOD data} \label{whenood}

For Naru/NeuroCard, DBEst++, and TVAE, and for each dataset, we compare 4 updating approaches against each other and against the base model before any new data is inserted. The 4 approaches are as follows: 
"\texttt{Retrain}", retrains the model from scratch using both old and new data.  "\texttt{Baseline}" is the baseline approach in Naru/NeuroCard and DBest++ where a trained model is updated with new data by performing \textit{SGD} with a smaller learning rate. "\texttt{DDUp}" is the proposed method. 
Finally, in "\texttt{stale}", the model is not updated -- this is a do-nothing approach.
For reference, we also include the numbers for $M_0$, i.e., the original model accuracy before any new data came.
\autoref{tab:qerror} and \autoref{tab:aqpacc} show the accuracy results for CE and AQP (SUM and AVG operations), respectively.
For TVAE, the classification f1-scores are reported in \autoref{tab:tvaef1}. Results of these three tables correspond to the case where the update sample is permuted. 
DDUp always performs better than the baseline approach. 
Most of the times, the performance of DBEst++ on DMV dataset is not as well as for the other datasets. This probably is due to the complexity of data (large scale and highly correlated attributes). Nevertheless, DDUp stands on the top of the underlying models and regardless of the model's performance, DDUp ensures that it will retain the accuracy.
Please note the DMV dataset results in \autoref{tab:qerror} and \autoref{tab:aqpacc} and, census and forest datasets in \autoref{tab:tvaef1}, where, DDUp even outperforms retraining from scratch. 
Interestingly, this corroborates similar evidence for sequential self-distillation (for boosting embeddings for) classification tasks \cite{seq-self-distill}. This was one of the reasons we adapted a self-distillation based approach.
Finally, baseline methods have poor performance for 95th and 99th percentiles. 

\begin{table*}[t]
  \caption{Results of updating a base model with a 20\% permuted sample in terms of q-error. $M_{0}$ denotes the base model.}
  \label{tab:qerror}
  \centering
  \begin{tabular}{c c | c | c | c | c | c | c | c | c | c | c} 
    \toprule
\multirow{2}{*}{Dataset} & \multirow{2}{*}{metric} & \multicolumn{5}{c|}{DBEst++} & \multicolumn{5}{c}{Naru/NeuroCard} \\
&&$M_{0}$&DDUp&baseline&stale&retrain&$M_{0}$&DDUp&baseline&stale&retrain \\
    \midrule
\multirow{4}{*}{census}&median&1.05&1.11&1.17&1.16&1.07&1.08&1.09&4&1.14&1.07\\
&95th&2&2&2.20&2&2&2&2&471.80&2&2\\
&99th&3&3&4&3&3&3&3&1534.69&3.16&3\\
&max&5&7&11&10.50&5&5.25&7&8385&21.88&6\\
\midrule
\multirow{4}{*}{forest}&median&1.026&1.046&2&1.18&1.02&1.04&1.07&1.54&1.10&1.05\\
&95th&2&2&63.40&2&1.64&2.48&3&41&2.50&2.75\\
&99th&2&2.583&503.12&5.60&2&4&6&157.16&5.48&5\\
&max&4&5.33&3470&90.85&5.33&27&65.66&1691&484&34.66\\
\midrule

\multirow{4}{*}{DMV}&median&1.20&1.143&3.48&1.88&1.34&1.02&1.04&2.57&1.16&1.02\\
&95th&4.91&5.07&234.88&7.00&5.50&1.20&1.41&468.68&1.50&1.25\\
&99th&9.65&10&3897.87&12.50&8&1.83&2.31&4734.62&2.84&2\\
&max&18.83&19&65875&39&17&8&9.81&343761&9.49&5\\

\midrule
\multirow{4}{*}{TPCDS}&median&1.02&1.04&57&1.27&1.02&1.01&1.07&1.15&1.10&1.05\\
&95th&1.16&1.26&269&1.58&1.18&2&2&29&2&2\\
&99th&1.5&1.61&1266&2.72&1.5&3.01&3.01&239&4&3\\
&max&3&3&4534&10.66&5.64&5&28&5100&28&24\\

    \bottomrule
  \end{tabular}
\end{table*}

\begin{table}[t]
  \caption{mean-relative-error for SUM and AVG aggregation functions for DBEst++.}
  \label{tab:aqpacc}
  \centering
  \resizebox{\linewidth}{!}{%
  \begin{tabular}{c c | c c c c c} 
    \toprule
Dataset&function&$M_{0}$&DDUp&baseline&stale&retrain\\
    \midrule
\multirow{2}{*}{census}&SUM&13.05&17.30&65.88&21.36&13.60\\
&AVG&1.89&2.36&8.15&2.37&1.97\\
\midrule
\multirow{2}{*}{forest}&SUM&10.11&15.51&88.73&24.59&10.14\\
&AVG&0.76&1.04&3.90&1.35&0.79\\
\midrule
\multirow{2}{*}{TPCDS}&SUM&4.53&6.37&61.40&22.64&5.12\\
&AVG&0.88&1.47&12&3.50&1.21\\
\midrule
\multirow{2}{*}{DMV}&SUM&76.73&85.29&423&97.00&110\\
&AVG&6.4&6.9&15.9&8.6&7.3\\

\bottomrule
  \end{tabular}}
\end{table}

\begin{table}[t]
  \caption{Classification results for TVAE in terms of micro f1. 'r' stands for real data, 's' stands for synthetic data.}
  \label{tab:tvaef1}
  \centering
  \resizebox{\linewidth}{!}{%
  \begin{tabular}{c | c c | c c | c c | c c | c c } 
    \toprule
\multirow{2}{*}{Dataset}
&\multicolumn{2}{c}{$M_{0}$}&\multicolumn{2}{c}{DDUp}&\multicolumn{2}{c}{baseline}&\multicolumn{2}{c}{stale}&\multicolumn{2}{c}{retrain}\\
&r&s&r&s&r&s&r&s&r&s\\
    \midrule
census&0.67&0.63&0.77&0.73&0.77&0.55&0.77&0.56&0.77&0.72\\
forest&0.84&0.69&0.89&0.78&0.89&0.63&0.89&0.60&0.89&0.74\\
DMV&0.97&0.97&0.98&0.97&0.98&0.92&0.98&0.93&0.98&0.98\\

\bottomrule
  \end{tabular}}
\end{table}

\subsubsection*{Performance on old and new queries} To better illustrate the effects of \catforget{} and \textit{intransigence} we elaborate on performance on FWT and BWT. (As \texttt{retrain} avoids be definition \catforget{} and \textit{intransigence}, it is omitted).
The results are shown in \autoref{tab:mdntranfers}. 
Note that any insertion affects only a percentage of queries, shown in 
\autoref{tab:querypercents}. 
Comparing AT, FWT, and BWT in \autoref{tab:qerror} and \autoref{tab:mdntranfers} first note that fine-tuning always performs much better in terms of FWT compared to BWT (due to catastrophic forgetting). 
Second, conversely, a stale model shows better BWT compared to FWT. 
For DDUp, FWT and BWT remain close to each other, especially in terms of median q-error, showing that DDUP can ensure accuracy for queries on old and new data. 
Overall, DDUp enjoys high accuracy.

\subsubsection*{Incremental Steps} 
\begin{figure}
\begin{minipage}{.5\linewidth}
\subfloat[]{\label{main:a}\includegraphics[scale=.29]{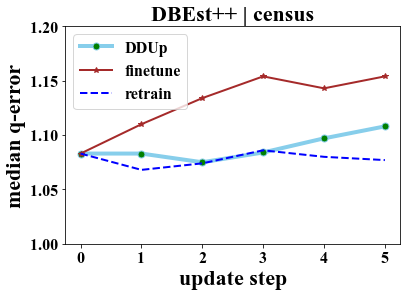}}
\end{minipage}%
\begin{minipage}{.5\linewidth}
\subfloat[]{\label{main:b}\includegraphics[scale=.29]{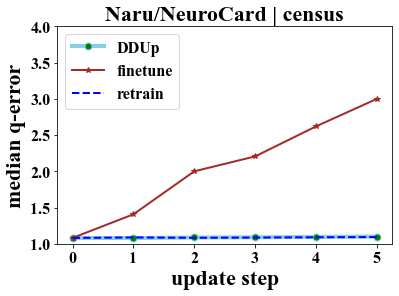}}
\end{minipage} %
\begin{minipage}{.5\linewidth}
\subfloat[]{\label{main:c}\includegraphics[scale=.29]{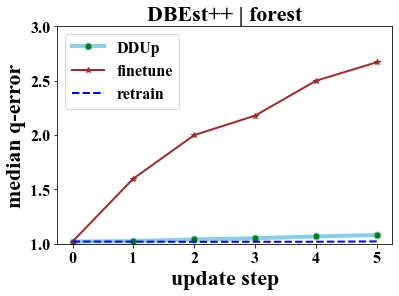}}
\end{minipage}%
\begin{minipage}{.5\linewidth}
\subfloat[]{\label{main:d}\includegraphics[scale=.29]{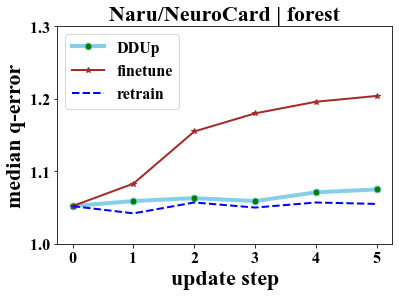}}
\end{minipage}
\begin{minipage}{.5\linewidth}
\subfloat[]{\label{main:e}\includegraphics[scale=.29]{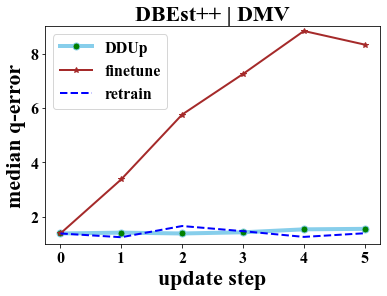}}
\end{minipage}%
\begin{minipage}{.5\linewidth}
\subfloat[]{\label{main:f}\includegraphics[scale=.29]{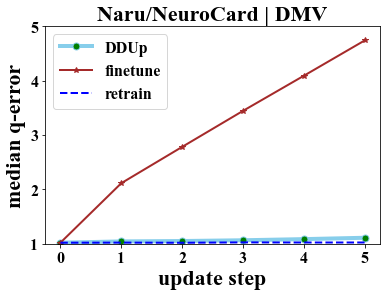}}
\end{minipage}
\vspace{-0.3cm}
\caption{Updating results over 5 consecutive updates.}
\label{fig:incupdates2}
\vspace{-0.4cm}
\end{figure}

To show the updates in incremental steps, we have split the \(20\%\) data into 5 equal-sized chunks and have performed an update incrementally for each batch. \autoref{fig:incupdates2} compares the trend of accuracy during updates. As it is clear from the figures, DDUp remains very close to \texttt{retrain}, while there is a drastic drop in accuracy using \texttt{baseline}. Starting point \(0\) is where the base model \(M_{0}\) is built from scratch. (The same results hold for 95th, 99th percentiles and maximum q-error).

\begin{table*}[t]
 \caption{Comparing q-error of different updating approaches in terms of FWT and BWT.}
 \vspace{-0.2cm}
  \label{tab:mdntranfers}
  \begin{tabular}{c c | c | c c | c c | c c | c | c c | c c | c c } 
    \toprule
\multirow{3}{*}{Dataset} & \multirow{3}{*}{metric} & \multicolumn{7}{c|}{DBEst++} & \multicolumn{7}{c}{Naru/NeuroCard} \\
&&\multicolumn{1}{c}{$M_{0}$}&\multicolumn{2}{c}{DDUp}&\multicolumn{2}{c}{baseline}&\multicolumn{2}{c|}{stale}&\multicolumn{1}{c}{$M_{0}$}&\multicolumn{2}{c}{DDUp}&\multicolumn{2}{c}{baseline}&\multicolumn{2}{c}{stale} \\

&&&FWT&BWT&FWT&BWT&FWT&BWT& &FWT&BWT&FWT&BWT&FWT&BWT\\

\midrule
\multirow{3}{*}{census}&median&1.05&1.06&1.12&1.06&1.20&1.05&1.16&1.08&1.11&1.09&1.83&6&1.20&1.13 \\
&95th&2&1.66&2&1.56&2.33&3.30&2&2&1.64&2&4.63&530.80&3.18&2 \\
&99th&3&4.94&3&4.10&4&8.90&2.75&3&3.08&3&9.98&1598.53&8.49&3\\

\midrule
\multirow{3}{*}{forest}&median &1.02&1.01&1.08&1.23&2.66&1.05&1.20&1.04&1.07&1.07&1.39&1.65&1.18&1.08\\
&95th&2&1.181&2&2.87&146.38&2.85&2&2.489&1.88&3&3.13&43.02&7.55&2.33\\
&99th&2&1.52&3&3.72&590.57&18.33&2.24&4&4.89&6&5.27&163.80&191.53&4.86\\

\midrule
\multirow{3}{*}{DMV}&median&1.20&1.28&1.13&2.20&4.36&1.66&1.54&1.02&1.02&1.07&1.06&12.85&1.26&1.19\\
&95th&4.910&4.30&5.87&3.34&484.46&9.50&6.87&1.20&1.16&1.55&1.65&1015.81&3.30&1.40\\
&99th&9.65&9&11.65&10.50&5894.21&12.12&10.80&1.83&1.47&3&3.35&8183.34&11.93&2.49\\

\midrule
\multirow{3}{*}{TPCDS}&median&1.02&1.03&1.04&1.20&1.51&1.16&1.21&1.01&1.06&1.08&1.19&1.11&1.10&1.10\\

&95th&1.16&1.21&1.29&2.37&339&2.26&1.35&2&2&2&2.60&54&2&2\\
&99th&1.5&1.37&1.66&4.27&1536&4.48&1.66&3.01&9.77&3&9.47&434&9.64&3.77\\

\bottomrule
  \end{tabular}
\end{table*}

\begin{table}[t]
  \caption{The percentage of the queries (out of 2k queries) with changed actual results after inserting 20\% new data.}
  \vspace{-0.2cm}
  \label{tab:querypercents}
  \begin{tabular}{c c c } 
    \toprule
     dataset & DBEst++ & Naru \\
     \midrule
     census&14\%&12\% \\
     forest&32\%&9\% \\
     TPCDS&36\%&36\% \\
     dmv&52\%&45\%\\
    \bottomrule
  \end{tabular}
  \vspace{-0.5cm}
\end{table}

\begin{figure}
\begin{minipage}{.5\linewidth}
\subfloat[]{\label{main:c}\includegraphics[scale=.29]{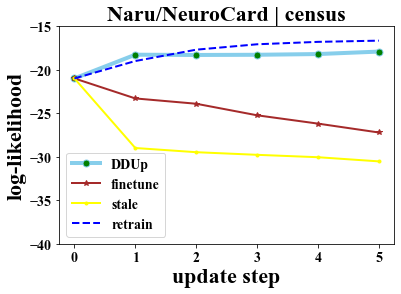}}
\end{minipage}%
\begin{minipage}{.5\linewidth}
\subfloat[]{\label{main:d}\includegraphics[scale=.29]{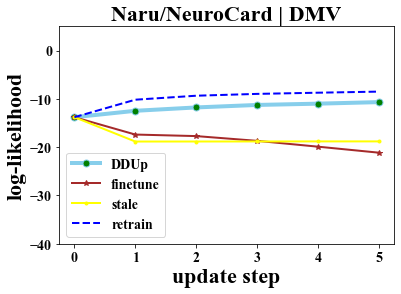}}
\end{minipage}
\vspace{-0.3cm}
\caption{log-likelihood results over 5 consecutive updates.}  
\label{fig:incll}
\vspace{-0.3cm}
\end{figure}

We also have evaluated the models with respect to the \textit{log-likelihood goodness-of-fit}. Log-likelihood is  widely used to evaluate NN models. 
Using log-likelihood allows evaluation to be independent 
of underlying applications. \autoref{fig:incll} shows changes in log-likelihood in consecutive update steps. At each step, we calculate the average of log-likelihoods over a sample of new data and a sample from historical data. In these figures we again see that updating with DDUp is fitting to the old and the new data very similarly to the \texttt{retrain} case.  In general, when keep using \texttt{stale}, the log-likelihood drops after the first update and then remains low. The reason is that all update batches have similar permutation and since we calculate unweighted averages, the log-likelihood stays fixed. While, for \texttt{baseline}, i.e fine-tuning, we can see a gradual decrease of likelihood which means that the network is increasingly forgetting about previous data in each step. 

\subsubsection{When data is not OOD}
\begin{figure}
\begin{minipage}{.5\linewidth}
\subfloat[]{\label{main:a}\includegraphics[scale=.29]{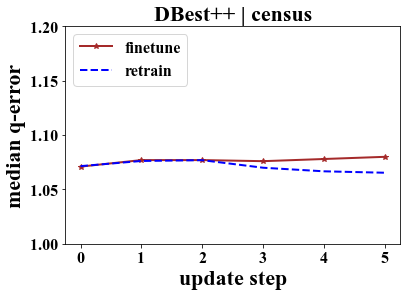}}
\end{minipage}%
\begin{minipage}{.5\linewidth}
\subfloat[]{\label{main:b}\includegraphics[scale=.29]{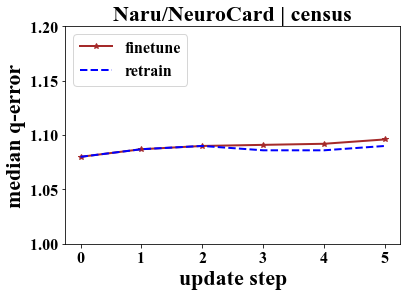}}
\end{minipage} %
\vspace{-0.3cm}
\caption{Updating results over 5 consecutive updates when data follows the same distribution as the historical data.}
\label{fig:incupdatenodrift}
\vspace{-0.3cm}
\end{figure}

In this case, simple fine-tuning update algorithms, such as \texttt{baseline}, will likely avoid \catforget{}. 
To illustrate this, we have repeated the 5 batched incremental updates with data without permutation. The results are reported in \autoref{fig:incupdatenodrift}. For space reasons, we only show the results for census. The results indicate that for in-distribution data, simple baselines can have a performance close to \texttt{retrain}.

\begin{table}[hb]
\vspace{-0.4cm}
  \caption{DDUp's speed up over \texttt{retrain}, for two update sizes. For census, forest, and dmv, sp1: 20\% of the original table. sp2, 5\% of the original table. for IMDB and TPCH sp1: updating the first partition and sp2: updating the last partition.}
  \label{tab:times}
\vspace{-0.25cm}
  \begin{tabular}{c | c c | c c | c c } 
    \toprule
\multirow{2}{*}{Dataset}  & \multicolumn{2}{c|}{DBEst++} & \multicolumn{2}{c|}{Naru} & \multicolumn{2}{c}{TVAE} \\
&sp1&sp2&sp1&sp2&sp1&sp2 \\
    \midrule
census&5&5.5&3.5&4&3.4&5.7 \\
forest&1.6&4&5&9.2&3.6&7 \\
DMV&4&6.5&2.3&9.6&3.4&6.8 \\
IMDB&4.5&18&3.5&5&NA&NA \\
tpch&6.5&16&2&4&NA&NA \\
    \bottomrule
  \end{tabular}
\end{table}

\subsection{Evaluating DDUp for Join Queries} \label{joinexp}
\begin{figure}[htbp]
\begin{minipage}{.5\linewidth}
\subfloat[]{\label{main:a}\includegraphics[scale=.29]{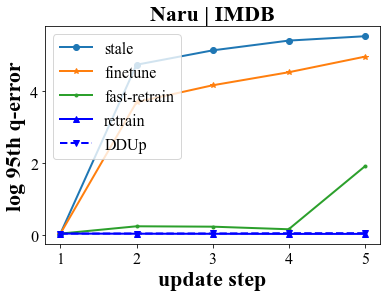}}
\end{minipage}%
\begin{minipage}{.5\linewidth}
\subfloat[]{\label{main:b}\includegraphics[scale=.29]{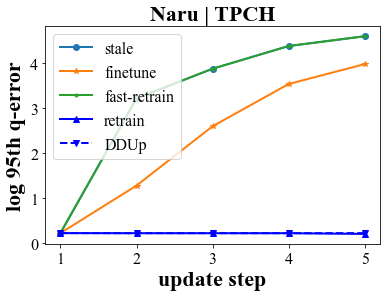}}
\end{minipage} \\
\begin{minipage}{.5\linewidth}
\subfloat[]{\label{main:c}\includegraphics[scale=.29]{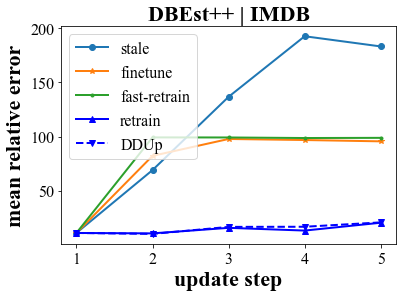}}
\end{minipage}%
\begin{minipage}{.5\linewidth}
\subfloat[]{\label{main:d}\includegraphics[scale=.29]{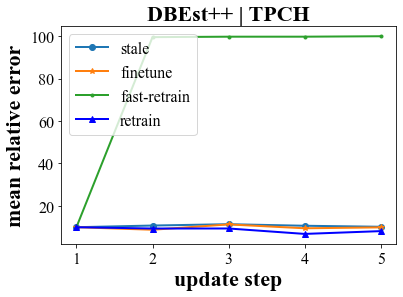}}
\end{minipage}
\vspace{-0.3cm}
\caption{DDUp's performance on joined tables.}
\label{fig:joins}
\vspace{-0.3cm}
\end{figure}

As mentioned, DDUp is unconcerned whether at a time $t$, 
\(S^{\leq}_{t-1}\) (a sample of  \(\cup_{j=0}^{t-1} D_j\)) and $D_t$ come from a raw table or from a join.
For this experiment, we have evaluated DDUp running 2,000 queries over two 3-table joins from the JOB and TPCH datasets.
For each, the 2,000 queries involve a join of the fact table with two dimension tables: 
Specifically, the join of tables [\texttt{title}, \texttt{movie info idx}, \texttt{movie companies}] for IMDB, and [\texttt{orders}, \texttt{customer}, \texttt{nation}] for TPCH. For the update dynamics, we have split the fact table into 5 time-ordered equally-sized partitions. We have built \(M_0\) on the join (of the fact table's first partition with the 2 dimension tables) and updated it with each subsequent partition at a time. This is similar to the update setting in NeuroCard. 
Results for both CE and AQP are in \autoref{fig:joins}.

NeuroCard, unlike other models, natively supports joins, using 
a "fast-retrain" - i.e., a light retraining where the model is retrained using a 1 percent sample of the full join result. We have included this policy here as "fast-retrain". 
DDUp always signalled OOD for the new update batches, except for TPCH data on DBest++, where update was not triggered. Therefore, in \autoref{fig:joins}.d the accuracy of the stale model and fine-tuning is  close to retrain. This further confirms the significance of OOD detection.

\subsection{Effect of Transfer Learning} \label{distilleval}
\begin{figure}[htbp]
    \centering
    \includegraphics[width=\linewidth]{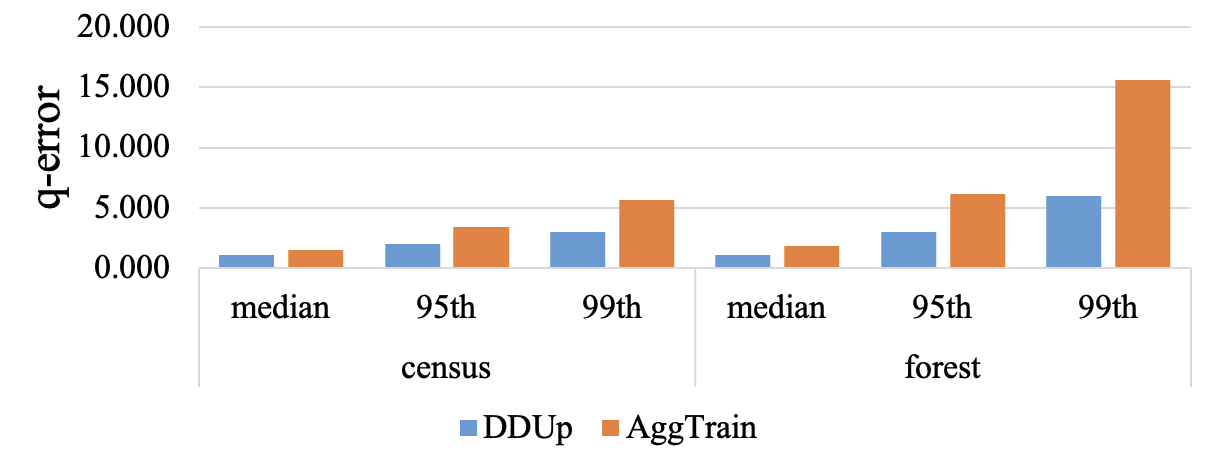}
    \caption{Effect of transfer-learning on q-error. AggTrain, is the case where we aggregate the transfer-set with the new data and train a model similar to the base model.}
    \label{fig:tleffect}
\end{figure}

We now delve into the effects of transfer-learning in DDUp. How much DDUp's transfer-learning via knowledge distillation contributes to better accuracy? 
We perform experiments where we remove the transfer-learning term of Eq \ref{totalloss}. Therefore, we combine the sample from previous data known as the transfer-set with the new update batch and create a model 
with the same configurations as the base model. \autoref{fig:tleffect} shows the results. 
The results assert that the performance of DDUp is not only related to the previous data sample, and in fact, distillation has a big effect on the improvement of the new models. 
\vspace{-0.2cm}
\subsection{Overheads} \label{overheads}
We report on the costs of each DDUp module separately. All the codes are written and executed in Python 3.8, on an Ubuntu 20 machine  with 40 CPU cores, two Nvidia GTX 2080 GPUs and 64GB memory. With respect to memory usage, DDUp performs regular feed-forward steps as in regular NN training. Therefore, DDUp does not increase memory footprints 
In terms of time, DDUp has two computation costs namely, \textit{OOD detection} and \textit{model update}. OOD detection is split into offline and online phases. \autoref{tab:offontime} shows these two times. The largest detection time is for the forest dataset on a Naru model which takes around 3 minutes. However, please note that in 
the online phase only takes 1 second to detect a change in data. 

\begin{table}[hb]
  \caption{online and offline times during OOD detection.}
  \label{tab:offontime}
\vspace{-0.25cm}
  \begin{tabular}{c | c c | c c | c c } 
    \toprule
\multirow{2}{*}{Dataset}  & \multicolumn{2}{c|}{DBEst++} & \multicolumn{2}{c|}{Naru} & \multicolumn{2}{c}{TVAE} \\
&off&on&off&on&off&on \\
    \midrule
census&2.44&0.02&111&1.8&310&5.5\\
forest&28&0.04&174&0.92&433&8.8\\
DMV&86&2&144&10&99&0.44\\
    \bottomrule
  \end{tabular}
\end{table}

\autoref{tab:times} shows DDUp's speed up over \texttt{retrain} for OOD data, for different update sizes. When data is OOD, DDUp can be over 9$\times$ faster than \texttt{retrain}. Obviously, speedups will be higher for incremental steps. This fact is reflected in IMDB and TPCH datasets where after inserting the last partition DDUp is 18$\times$ faster than \texttt{retrain}. Note that the updating time is dependent on a few parameters including update size, transfer-set size, training batch size etc. During updates, we have used smaller training batch sizes. If one tunes the model for bigger batches, and smaller transfer-set sizes, the speed up would be higher.

\vspace{-0.2cm}
\subsection{Non neural network models}\label{nonnn}
For the sake of completeness and as an additional reference point, we include results for updating a state-of-the-art non-NN model that natively supports data insertions, (DeepDB \cite{hilprecht2019deepdb}) used for CE. 
When an update happens, DeepDB traverses its sum-product-network graph and updates the weights of the intermediate nodes and the histograms at the leaves. We have repeated the same experiment in \autoref{tab:qerror} for DeepDB. The results are reported in \autoref{tab:deepdb}.

\begin{table}[t]
  \caption{Performance of DeepDB updating vs. DDUp for Naru, for a CE task in terms of q-error.}
    \vspace{-0.35cm}
  \label{tab:deepdb}
  \centering
  \begin{tabular}{c c | c | c | c | c | c } 
    \toprule
\multirow{2}{*}{Dataset} & \multirow{2}{*}{ metric} & \multicolumn{3}{c|}{DeepDB} & \multicolumn{2}{c}{Naru} \\
&&$M_{0}$&update&retrain&$M_{0}$&DDUp \\
    \midrule
\multirow{3}{*}{census}&median&1.05&1.2&1.05&1.08&1.09\\
&95th&3&4.18&3&2&2\\
&99th&5.11&8&5&3&3\\
\midrule
\multirow{3}{*}{forest}&median&1.02&1.2&1.02&1.04&1.07\\
&95th&7.5&10.5&7&2.48&3\\
&99th&31&52&31&4&6\\
\midrule
\multirow{3}{*}{DMV}&median&1.06&1.25&1.1&1.02&1.04\\
&95th&2.5&3.5&2.5&1.20&1.41\\
&99th&22&37&21&1.83&2.31\\
    \bottomrule
  \end{tabular}
  \vspace{-0.3cm}
\end{table}

From \autoref{tab:deepdb} it can be observed that DeepDB's updating policy is under-performing, as was independently verified in \cite{wang2020we}. 
DDUp (coupled in this experiment with Naru/NeuroCard for CE) always performs better. Nonetheless, we wish to emphasize that the saving grace for DeepDB based on our experiments is that retraining from scratch is very efficient -- significantly faster compared to NNs. 

\section{Related Work} \label{litraturere}
\subsection{Learned Database Systems}\label{ldbliterature}
NN-based components to be used by DBs are emerging rapidly. Different works exploit different neural network models. 
\cite{yang2019deep, yang2020neurocard, hasan2020deep} used generative neural networks to build learned selectivity estimators. Thirumuruganathan et al. \cite{thirumuruganathan2020approximate} used VAEs for AQP. Ma et al. \cite{ma2021learned} used mixture density networks for AQP. Database indexing research 
recently has adopted neural networks to approximate cumulative density functions \cite{kraska2018case,ding2020alex,nathan2020learning,ding2020tsunami}. Query optimization and join ordering are also benefiting from neural networks \cite{marcus2019neo, kipf2018learned}. Other applications include auto-tuning databases \cite{van2017automatic,li2019qtune,zhang2019end}, cost estimation \cite{zhi2021efficient, siddiqui2020cost}, and workload forecasting \cite{zhu2019novel}.

Among these, this work provides a solution for handling NN model maintenance in the face of  insertion-updates with OOD data, when the models need to continue ensuring high accuracy on new and old data and on tasks for which models were originally trained (such as AQP, CE, DG, etc.).
While there has been related research on transfer learning for learned DBs such as \cite{hilprecht2021one, wu2021unified} these target a different problem setting:
They study how to transfer knowledge from a model trained for one task, and/or a DB, and/or a system, and/or a workload to a new task and/or DB, and/or system, and/or workload. They do not study how to keep performing the original task(s) on evolving datasets with insertions carrying OOD data with high accuracy for queries on both old and new data. Simply using these methods by fine-tuning on new data will incur catastrophic forgetting. Nevertheless, since these models employ some sorts of knowledge transfer, they might be useful to support updates. However, it remains open whether and how the models in \cite{wu2021unified, hilprecht2021one} can be utilized to solve efficiently the problems tackled in this paper.
While some of non-neural-network models (e.g., DeepDB) can very efficiently retrain from scratch, 
NN-based models for the above problem setting either do not support insertion-updates or suffer from poor accuracy when facing OOD data, unless paying the high costs of retraining from scratch.

\subsection{OOD Detection}
OOD detection has recently attracted a lot of attention and it has long been studied in statistics as concept drift (CD) detection, or novelty detection. In general, CD and OOD detection methods could be divided into two broad categories \cite{gama2014survey,lu2018learning,wang2020few}: 
First, prediction-based methods, which use the predictions of the underlying models to test for a change. Recent ML models usually use the predictive probabilities of the classifiers as a confidence score to identify changes \cite{jiang2018trust,ovadia2019can, wilson2020bayesian,ruff2021unifying}. Others may monitor the error of the underlying models and trigger an OOD signal when a significant change is captured \cite{gama2006learning, baena2006early,savva2019aggregate,nehme2009self,lopez2016revisiting}. While these approaches are very efficient in time, they typically come with limiting assumptions depending on the underlying model or application. For example, most of them can only be utilized and are only studied for classification (supervised) tasks.
The second broad family of methods is distribution-based methods. Some of these methods try to find a distance measure that can best show the discrepancy between new data and old data distributions, using tests like Kolmogorov-Smirnov (KS), \cite{kolmogorov1933sulla}, Wilcoxon \cite{pereira2009machine}, and their multi-variate variants \cite{fasano1987multidimensional, baringhaus2004new}. Others try to learn the density of the underlying data distribution test for a significant change, like kernel-density-based approaches  \cite{kifer2004detecting,dasu2006information,gu2016concept,lu2014concept,bu2016pdf,song2007statistical}. More recent works utilize the estimated likelihoods of generative models \cite{ren2019likelihood, morningstar2021density, xiao2020likelihood}. Other approaches rely on the inner representations of the networks \cite{li2021cutpaste,hendrycks2019using,lee2018simple}. Nonetheless, this second family of OOD detection methods are usually expensive (esp. for multi-dimensional data) and involve fitting a separate density estimator. Hence, the main problem is that in an insertion scenario, the density estimators also need to be updated (typically via training from scratch, upon each insertion).

\vspace{-0.3cm}
\subsection{Incremental Learning (IL)}
Most IL methods regularize the model in a way that it acquires knowledge from the new task while retaining the knowledge of old tasks. For example, \textit{Elastic Weight Consolidation (EWC)} \cite{kirkpatrick2017overcoming} adds a regularizer to control the learning speed around important weights of the network for old tasks while learning a new task. Similar works are developed around this idea \cite{liu2018rotate, lee2020continual,titsias2019functional}, \textit{Path Integral (PathInt)} \cite{zenke2017continual} ,\textit{Riemanian Walk (RWalk)} \cite{chaudhry2018riemannian}. Other approaches exploit knowledge distillation to retain the knowledge of previous tasks \cite{li2017learning}.
Another group of IL methods, save exemplars from past data \cite{wu2019large, castro2018end, rebuffi2017icarl} or generate samples/features using generative models \cite{ostapenko2019learning, kemker2017fearnet} and involve them in learning new tasks. Lopez et al. \cite{lopez2017gradient} has proposed \textit{Gradient Episodic Memory} that consists of \textit{M} blocks of memory to store examples from \textit{T} tasks and uses the model's prediction on these examples as a constraining loss that inhibits the model to bias toward new task and forget past tasks. Lastly, some works try to completely keep previous models and create new models (or part of a model like a single layer) for each new task.  Aljundi et al. \cite{aljundi2017expert} introduce \textit{Expert Gate} with different models for each task and an autoencoder which learns the representations of each task to assign test-time tasks to the proper model. Instead of learning a whole new model, Rusu et al. \cite{rusu2016progressive} introduce \textit{Progressing Neural Networks} which add new columns to the previous network architecture and learns lateral connections between them. Most of the above methods, do not account for in- and out- of distribution updates and are not easily extendable to different learning tasks. 

\vspace{-0.2cm}
\section{Conclusion} \label{conclusion}
Learned DB components can become highly  inaccurate when faced with new OOD data when aiming to ensure high accuracy for queries on old and new data for their original learning tasks. 
This work proposes, to our knowledge, the first solution to this problem, coined DDUp.
DDUp entails two novel components, for OOD detection and model updating.
To make detection widely applicable, OOD detection in DDUp exploits the output of the neural network (be it based on log-likelihood, cross-entropy, ELBO loss, etc.), and utilizes a principled two-sample test and a bootstrapping method to efficiently derive and use thresholds to signal OOD data.
DDUp also offers a general solution for model updating based on sequential self-distillation and a new loss function which carefully accounts for \textit{catastrophic forgetting} and \textit{intransigence}.
This work showcases the wide applicability of DDUp model updating by instantiating the general approach to three important learned functions for data management, namely AQP, CE, and DG, whereby a different type of NN (MDNs, DARNs, VAEs) is used for each. In fact, to our knowledge, no prior work has shown how to "distill-and-update" MDNs, VAEs, and DARNs.
Comprehensive experimentation showcases that DDUp detects OOD accurately and ensures high accuracy with its updated models with very low overheads. 


\section{Acknowledgement}
This work is partially sponsored by Huawei IRC and by EPSRC while doing a PhD at the University of Warwick.

\balance

\bibliographystyle{ACM-Reference-Format}
\bibliography{arxiv}


\begin{thebibliography}{85}


\ifx \showCODEN    \undefined \def \showCODEN     #1{\unskip}     \fi
\ifx \showDOI      \undefined \def \showDOI       #1{#1}\fi
\ifx \showISBNx    \undefined \def \showISBNx     #1{\unskip}     \fi
\ifx \showISBNxiii \undefined \def \showISBNxiii  #1{\unskip}     \fi
\ifx \showISSN     \undefined \def \showISSN      #1{\unskip}     \fi
\ifx \showLCCN     \undefined \def \showLCCN      #1{\unskip}     \fi
\ifx \shownote     \undefined \def \shownote      #1{#1}          \fi
\ifx \showarticletitle \undefined \def \showarticletitle #1{#1}   \fi
\ifx \showURL      \undefined \def \showURL       {\relax}        \fi
\providecommand\bibfield[2]{#2}
\providecommand\bibinfo[2]{#2}
\providecommand\natexlab[1]{#1}
\providecommand\showeprint[2][]{arXiv:#2}

\bibitem[Aljundi et~al\mbox{.}(2017)]%
        {aljundi2017expert}
\bibfield{author}{\bibinfo{person}{Rahaf Aljundi}, \bibinfo{person}{Punarjay
  Chakravarty}, {and} \bibinfo{person}{Tinne Tuytelaars}.}
  \bibinfo{year}{2017}\natexlab{}.
\newblock \showarticletitle{Expert gate: Lifelong learning with a network of
  experts}. In \bibinfo{booktitle}{\emph{Proceedings of the IEEE Conference on
  Computer Vision and Pattern Recognition}}. \bibinfo{pages}{3366--3375}.
\newblock


\bibitem[Baena-Garc{\i}a et~al\mbox{.}(2006)]%
        {baena2006early}
\bibfield{author}{\bibinfo{person}{Manuel Baena-Garc{\i}a},
  \bibinfo{person}{Jos{\'e} del Campo-{\'A}vila}, \bibinfo{person}{Ra{\'u}l
  Fidalgo}, \bibinfo{person}{Albert Bifet}, \bibinfo{person}{R Gavalda}, {and}
  \bibinfo{person}{Rafael Morales-Bueno}.} \bibinfo{year}{2006}\natexlab{}.
\newblock \showarticletitle{Early drift detection method}. In
  \bibinfo{booktitle}{\emph{Fourth international workshop on knowledge
  discovery from data streams}}, Vol.~\bibinfo{volume}{6}.
  \bibinfo{pages}{77--86}.
\newblock


\bibitem[Baringhaus and Franz(2004)]%
        {baringhaus2004new}
\bibfield{author}{\bibinfo{person}{Ludwig Baringhaus} {and}
  \bibinfo{person}{Carsten Franz}.} \bibinfo{year}{2004}\natexlab{}.
\newblock \showarticletitle{On a new multivariate two-sample test}.
\newblock \bibinfo{journal}{\emph{Journal of multivariate analysis}}
  \bibinfo{volume}{88}, \bibinfo{number}{1} (\bibinfo{year}{2004}),
  \bibinfo{pages}{190--206}.
\newblock


\bibitem[Bu et~al\mbox{.}(2016)]%
        {bu2016pdf}
\bibfield{author}{\bibinfo{person}{Li Bu}, \bibinfo{person}{Cesare Alippi},
  {and} \bibinfo{person}{Dongbin Zhao}.} \bibinfo{year}{2016}\natexlab{}.
\newblock \showarticletitle{A pdf-free change detection test based on density
  difference estimation}.
\newblock \bibinfo{journal}{\emph{IEEE transactions on neural networks and
  learning systems}} \bibinfo{volume}{29}, \bibinfo{number}{2}
  (\bibinfo{year}{2016}), \bibinfo{pages}{324--334}.
\newblock


\bibitem[Castro et~al\mbox{.}(2018)]%
        {castro2018end}
\bibfield{author}{\bibinfo{person}{Francisco~M Castro},
  \bibinfo{person}{Manuel~J Mar{\'\i}n-Jim{\'e}nez},
  \bibinfo{person}{Nicol{\'a}s Guil}, \bibinfo{person}{Cordelia Schmid}, {and}
  \bibinfo{person}{Karteek Alahari}.} \bibinfo{year}{2018}\natexlab{}.
\newblock \showarticletitle{End-to-end incremental learning}. In
  \bibinfo{booktitle}{\emph{Proceedings of the European conference on computer
  vision (ECCV)}}. \bibinfo{pages}{233--248}.
\newblock


\bibitem[Chaudhry et~al\mbox{.}(2018)]%
        {chaudhry2018riemannian}
\bibfield{author}{\bibinfo{person}{Arslan Chaudhry}, \bibinfo{person}{Puneet~K
  Dokania}, \bibinfo{person}{Thalaiyasingam Ajanthan}, {and}
  \bibinfo{person}{Philip~HS Torr}.} \bibinfo{year}{2018}\natexlab{}.
\newblock \showarticletitle{Riemannian walk for incremental learning:
  Understanding forgetting and intransigence}. In
  \bibinfo{booktitle}{\emph{Proceedings of the European Conference on Computer
  Vision (ECCV)}}. \bibinfo{pages}{532--547}.
\newblock


\bibitem[Choi et~al\mbox{.}(2017)]%
        {choi2017generating}
\bibfield{author}{\bibinfo{person}{Edward Choi}, \bibinfo{person}{Siddharth
  Biswal}, \bibinfo{person}{Bradley Malin}, \bibinfo{person}{Jon Duke},
  \bibinfo{person}{Walter~F Stewart}, {and} \bibinfo{person}{Jimeng Sun}.}
  \bibinfo{year}{2017}\natexlab{}.
\newblock \showarticletitle{Generating multi-label discrete patient records
  using generative adversarial networks}. In \bibinfo{booktitle}{\emph{Machine
  learning for healthcare conference}}. PMLR, \bibinfo{pages}{286--305}.
\newblock


\bibitem[Dasu et~al\mbox{.}(2006)]%
        {dasu2006information}
\bibfield{author}{\bibinfo{person}{Tamraparni Dasu}, \bibinfo{person}{Shankar
  Krishnan}, \bibinfo{person}{Suresh Venkatasubramanian}, {and}
  \bibinfo{person}{Ke Yi}.} \bibinfo{year}{2006}\natexlab{}.
\newblock \showarticletitle{An information-theoretic approach to detecting
  changes in multi-dimensional data streams}. In \bibinfo{booktitle}{\emph{In
  Proc. Symp. on the Interface of Statistics, Computing Science, and
  Applications}}. Citeseer.
\newblock


\bibitem[Ding et~al\mbox{.}(2020a)]%
        {ding2020alex}
\bibfield{author}{\bibinfo{person}{Jialin Ding}, \bibinfo{person}{Umar~Farooq
  Minhas}, \bibinfo{person}{Jia Yu}, \bibinfo{person}{Chi Wang},
  \bibinfo{person}{Jaeyoung Do}, \bibinfo{person}{Yinan Li},
  \bibinfo{person}{Hantian Zhang}, \bibinfo{person}{Badrish Chandramouli},
  \bibinfo{person}{Johannes Gehrke}, \bibinfo{person}{Donald Kossmann},
  {et~al\mbox{.}}} \bibinfo{year}{2020}\natexlab{a}.
\newblock \showarticletitle{ALEX: an updatable adaptive learned index}. In
  \bibinfo{booktitle}{\emph{Proceedings of the 2020 ACM SIGMOD International
  Conference on Management of Data}}. \bibinfo{pages}{969--984}.
\newblock


\bibitem[Ding et~al\mbox{.}(2020b)]%
        {ding2020tsunami}
\bibfield{author}{\bibinfo{person}{Jialin Ding}, \bibinfo{person}{Vikram
  Nathan}, \bibinfo{person}{Mohammad Alizadeh}, {and} \bibinfo{person}{Tim
  Kraska}.} \bibinfo{year}{2020}\natexlab{b}.
\newblock \showarticletitle{Tsunami: A learned multi-dimensional index for
  correlated data and skewed workloads}.
\newblock \bibinfo{journal}{\emph{arXiv preprint arXiv:2006.13282}}
  (\bibinfo{year}{2020}).
\newblock


\bibitem[Fasano and Franceschini(1987)]%
        {fasano1987multidimensional}
\bibfield{author}{\bibinfo{person}{Giovanni Fasano} {and}
  \bibinfo{person}{Alberto Franceschini}.} \bibinfo{year}{1987}\natexlab{}.
\newblock \showarticletitle{A multidimensional version of the
  Kolmogorov--Smirnov test}.
\newblock \bibinfo{journal}{\emph{Monthly Notices of the Royal Astronomical
  Society}} \bibinfo{volume}{225}, \bibinfo{number}{1} (\bibinfo{year}{1987}),
  \bibinfo{pages}{155--170}.
\newblock


\bibitem[Furlanello et~al\mbox{.}(2018)]%
        {furlanello2018born}
\bibfield{author}{\bibinfo{person}{Tommaso Furlanello},
  \bibinfo{person}{Zachary Lipton}, \bibinfo{person}{Michael Tschannen},
  \bibinfo{person}{Laurent Itti}, {and} \bibinfo{person}{Anima Anandkumar}.}
  \bibinfo{year}{2018}\natexlab{}.
\newblock \showarticletitle{Born again neural networks}. In
  \bibinfo{booktitle}{\emph{International Conference on Machine Learning}}.
  PMLR, \bibinfo{pages}{1607--1616}.
\newblock


\bibitem[Gama and Castillo(2006)]%
        {gama2006learning}
\bibfield{author}{\bibinfo{person}{Joao Gama} {and} \bibinfo{person}{Gladys
  Castillo}.} \bibinfo{year}{2006}\natexlab{}.
\newblock \showarticletitle{Learning with local drift detection}. In
  \bibinfo{booktitle}{\emph{International conference on advanced data mining
  and applications}}. Springer, \bibinfo{pages}{42--55}.
\newblock


\bibitem[Gama et~al\mbox{.}(2014)]%
        {gama2014survey}
\bibfield{author}{\bibinfo{person}{Jo{\~a}o Gama}, \bibinfo{person}{Indr{\.e}
  {\v{Z}}liobait{\.e}}, \bibinfo{person}{Albert Bifet}, \bibinfo{person}{Mykola
  Pechenizkiy}, {and} \bibinfo{person}{Abdelhamid Bouchachia}.}
  \bibinfo{year}{2014}\natexlab{}.
\newblock \showarticletitle{A survey on concept drift adaptation}.
\newblock \bibinfo{journal}{\emph{ACM computing surveys (CSUR)}}
  \bibinfo{volume}{46}, \bibinfo{number}{4} (\bibinfo{year}{2014}),
  \bibinfo{pages}{1--37}.
\newblock


\bibitem[Golatkar et~al\mbox{.}(2020)]%
        {golatkar2020eternal}
\bibfield{author}{\bibinfo{person}{Aditya Golatkar},
  \bibinfo{person}{Alessandro Achille}, {and} \bibinfo{person}{Stefano
  Soatto}.} \bibinfo{year}{2020}\natexlab{}.
\newblock \showarticletitle{Eternal sunshine of the spotless net: Selective
  forgetting in deep networks}. In \bibinfo{booktitle}{\emph{Proceedings of the
  IEEE/CVF Conference on Computer Vision and Pattern Recognition}}.
  \bibinfo{pages}{9304--9312}.
\newblock


\bibitem[Gu et~al\mbox{.}(2016)]%
        {gu2016concept}
\bibfield{author}{\bibinfo{person}{Feng Gu}, \bibinfo{person}{Guangquan Zhang},
  \bibinfo{person}{Jie Lu}, {and} \bibinfo{person}{Chin-Teng Lin}.}
  \bibinfo{year}{2016}\natexlab{}.
\newblock \showarticletitle{Concept drift detection based on equal density
  estimation}. In \bibinfo{booktitle}{\emph{2016 International Joint Conference
  on Neural Networks (IJCNN)}}. IEEE, \bibinfo{pages}{24--30}.
\newblock


\bibitem[Hasan et~al\mbox{.}(2020)]%
        {hasan2020deep}
\bibfield{author}{\bibinfo{person}{Shohedul Hasan}, \bibinfo{person}{Saravanan
  Thirumuruganathan}, \bibinfo{person}{Jees Augustine}, \bibinfo{person}{Nick
  Koudas}, {and} \bibinfo{person}{Gautam Das}.}
  \bibinfo{year}{2020}\natexlab{}.
\newblock \showarticletitle{Deep learning models for selectivity estimation of
  multi-attribute queries}. In \bibinfo{booktitle}{\emph{Proceedings of the
  2020 ACM SIGMOD International Conference on Management of Data}}.
  \bibinfo{pages}{1035--1050}.
\newblock


\bibitem[Hendrycks and Gimpel(2017)]%
        {detectOOD-iclr17}
\bibfield{author}{\bibinfo{person}{Dan Hendrycks} {and} \bibinfo{person}{Kevin
  Gimpel}.} \bibinfo{year}{2017}\natexlab{}.
\newblock \showarticletitle{A Baseline for Detecting Misclassified and
  Out-of-Distribution Examples in Neural Networks.}. In
  \bibinfo{booktitle}{\emph{ICLR}}.
\newblock


\bibitem[Hendrycks et~al\mbox{.}(2019)]%
        {hendrycks2019using}
\bibfield{author}{\bibinfo{person}{Dan Hendrycks}, \bibinfo{person}{Kimin Lee},
  {and} \bibinfo{person}{Mantas Mazeika}.} \bibinfo{year}{2019}\natexlab{}.
\newblock \showarticletitle{Using pre-training can improve model robustness and
  uncertainty}. In \bibinfo{booktitle}{\emph{International Conference on
  Machine Learning}}. PMLR, \bibinfo{pages}{2712--2721}.
\newblock


\bibitem[Hilprecht and Binnig(2021)]%
        {hilprecht2021one}
\bibfield{author}{\bibinfo{person}{Benjamin Hilprecht} {and}
  \bibinfo{person}{Carsten Binnig}.} \bibinfo{year}{2021}\natexlab{}.
\newblock \showarticletitle{One Model to Rule them All: Towards Zero-Shot
  Learning for Databases}.
\newblock \bibinfo{journal}{\emph{arXiv preprint arXiv:2105.00642}}
  (\bibinfo{year}{2021}).
\newblock


\bibitem[Hilprecht et~al\mbox{.}(2019)]%
        {hilprecht2019deepdb}
\bibfield{author}{\bibinfo{person}{Benjamin Hilprecht},
  \bibinfo{person}{Andreas Schmidt}, \bibinfo{person}{Moritz Kulessa},
  \bibinfo{person}{Alejandro Molina}, \bibinfo{person}{Kristian Kersting},
  {and} \bibinfo{person}{Carsten Binnig}.} \bibinfo{year}{2019}\natexlab{}.
\newblock \showarticletitle{Deepdb: Learn from data, not from queries!}
\newblock \bibinfo{journal}{\emph{arXiv preprint arXiv:1909.00607}}
  (\bibinfo{year}{2019}).
\newblock


\bibitem[Hinton et~al\mbox{.}(2015)]%
        {hinton2015distilling}
\bibfield{author}{\bibinfo{person}{Geoffrey Hinton}, \bibinfo{person}{Oriol
  Vinyals}, {and} \bibinfo{person}{Jeff Dean}.}
  \bibinfo{year}{2015}\natexlab{}.
\newblock \showarticletitle{Distilling the knowledge in a neural network}.
\newblock \bibinfo{journal}{\emph{arXiv preprint arXiv:1503.02531}}
  (\bibinfo{year}{2015}).
\newblock


\bibitem[Jiang et~al\mbox{.}(2018)]%
        {jiang2018trust}
\bibfield{author}{\bibinfo{person}{Heinrich Jiang}, \bibinfo{person}{Been Kim},
  \bibinfo{person}{Melody Guan}, {and} \bibinfo{person}{Maya Gupta}.}
  \bibinfo{year}{2018}\natexlab{}.
\newblock \showarticletitle{To trust or not to trust a classifier}.
\newblock \bibinfo{journal}{\emph{Advances in neural information processing
  systems}}  \bibinfo{volume}{31} (\bibinfo{year}{2018}).
\newblock


\bibitem[Kemker and Kanan(2017)]%
        {kemker2017fearnet}
\bibfield{author}{\bibinfo{person}{Ronald Kemker} {and}
  \bibinfo{person}{Christopher Kanan}.} \bibinfo{year}{2017}\natexlab{}.
\newblock \showarticletitle{Fearnet: Brain-inspired model for incremental
  learning}.
\newblock \bibinfo{journal}{\emph{arXiv preprint arXiv:1711.10563}}
  (\bibinfo{year}{2017}).
\newblock


\bibitem[Kifer et~al\mbox{.}(2004)]%
        {kifer2004detecting}
\bibfield{author}{\bibinfo{person}{Daniel Kifer}, \bibinfo{person}{Shai
  Ben-David}, {and} \bibinfo{person}{Johannes Gehrke}.}
  \bibinfo{year}{2004}\natexlab{}.
\newblock \showarticletitle{Detecting change in data streams}. In
  \bibinfo{booktitle}{\emph{VLDB}}, Vol.~\bibinfo{volume}{4}. Toronto, Canada,
  \bibinfo{pages}{180--191}.
\newblock


\bibitem[Kingma and Welling(2013)]%
        {kingma2013auto}
\bibfield{author}{\bibinfo{person}{Diederik~P Kingma} {and}
  \bibinfo{person}{Max Welling}.} \bibinfo{year}{2013}\natexlab{}.
\newblock \showarticletitle{Auto-encoding variational bayes}.
\newblock \bibinfo{journal}{\emph{arXiv preprint arXiv:1312.6114}}
  (\bibinfo{year}{2013}).
\newblock


\bibitem[Kipf et~al\mbox{.}(2018)]%
        {kipf2018learned}
\bibfield{author}{\bibinfo{person}{Andreas Kipf}, \bibinfo{person}{Thomas
  Kipf}, \bibinfo{person}{Bernhard Radke}, \bibinfo{person}{Viktor Leis},
  \bibinfo{person}{Peter Boncz}, {and} \bibinfo{person}{Alfons Kemper}.}
  \bibinfo{year}{2018}\natexlab{}.
\newblock \showarticletitle{Learned cardinalities: Estimating correlated joins
  with deep learning}.
\newblock \bibinfo{journal}{\emph{arXiv preprint arXiv:1809.00677}}
  (\bibinfo{year}{2018}).
\newblock


\bibitem[Kirkpatrick et~al\mbox{.}(2017)]%
        {kirkpatrick2017overcoming}
\bibfield{author}{\bibinfo{person}{James Kirkpatrick}, \bibinfo{person}{Razvan
  Pascanu}, \bibinfo{person}{Neil Rabinowitz}, \bibinfo{person}{Joel Veness},
  \bibinfo{person}{Guillaume Desjardins}, \bibinfo{person}{Andrei~A Rusu},
  \bibinfo{person}{Kieran Milan}, \bibinfo{person}{John Quan},
  \bibinfo{person}{Tiago Ramalho}, \bibinfo{person}{Agnieszka
  Grabska-Barwinska}, {et~al\mbox{.}}} \bibinfo{year}{2017}\natexlab{}.
\newblock \showarticletitle{Overcoming catastrophic forgetting in neural
  networks}.
\newblock \bibinfo{journal}{\emph{Proceedings of the national academy of
  sciences}} \bibinfo{volume}{114}, \bibinfo{number}{13}
  (\bibinfo{year}{2017}), \bibinfo{pages}{3521--3526}.
\newblock


\bibitem[Kolmogorov(1933)]%
        {kolmogorov1933sulla}
\bibfield{author}{\bibinfo{person}{Andrey Kolmogorov}.}
  \bibinfo{year}{1933}\natexlab{}.
\newblock \showarticletitle{Sulla determinazione empirica di una lgge di
  distribuzione}.
\newblock \bibinfo{journal}{\emph{Inst. Ital. Attuari, Giorn.}}
  \bibinfo{volume}{4} (\bibinfo{year}{1933}), \bibinfo{pages}{83--91}.
\newblock


\bibitem[Kraska et~al\mbox{.}(2018)]%
        {kraska2018case}
\bibfield{author}{\bibinfo{person}{Tim Kraska}, \bibinfo{person}{Alex Beutel},
  \bibinfo{person}{Ed~H Chi}, \bibinfo{person}{Jeffrey Dean}, {and}
  \bibinfo{person}{Neoklis Polyzotis}.} \bibinfo{year}{2018}\natexlab{}.
\newblock \showarticletitle{The case for learned index structures}. In
  \bibinfo{booktitle}{\emph{Proceedings of the 2018 International Conference on
  Management of Data}}. \bibinfo{pages}{489--504}.
\newblock


\bibitem[Lee et~al\mbox{.}(2020)]%
        {lee2020continual}
\bibfield{author}{\bibinfo{person}{Janghyeon Lee}, \bibinfo{person}{Hyeong~Gwon
  Hong}, \bibinfo{person}{Donggyu Joo}, {and} \bibinfo{person}{Junmo Kim}.}
  \bibinfo{year}{2020}\natexlab{}.
\newblock \showarticletitle{Continual learning with extended kronecker-factored
  approximate curvature}. In \bibinfo{booktitle}{\emph{Proceedings of the
  IEEE/CVF Conference on Computer Vision and Pattern Recognition}}.
  \bibinfo{pages}{9001--9010}.
\newblock


\bibitem[Lee et~al\mbox{.}(2018)]%
        {lee2018simple}
\bibfield{author}{\bibinfo{person}{Kimin Lee}, \bibinfo{person}{Kibok Lee},
  \bibinfo{person}{Honglak Lee}, {and} \bibinfo{person}{Jinwoo Shin}.}
  \bibinfo{year}{2018}\natexlab{}.
\newblock \showarticletitle{A simple unified framework for detecting
  out-of-distribution samples and adversarial attacks}.
\newblock \bibinfo{journal}{\emph{Advances in neural information processing
  systems}}  \bibinfo{volume}{31} (\bibinfo{year}{2018}).
\newblock


\bibitem[Li et~al\mbox{.}(2021)]%
        {li2021cutpaste}
\bibfield{author}{\bibinfo{person}{Chun-Liang Li}, \bibinfo{person}{Kihyuk
  Sohn}, \bibinfo{person}{Jinsung Yoon}, {and} \bibinfo{person}{Tomas
  Pfister}.} \bibinfo{year}{2021}\natexlab{}.
\newblock \showarticletitle{Cutpaste: Self-supervised learning for anomaly
  detection and localization}. In \bibinfo{booktitle}{\emph{Proceedings of the
  IEEE/CVF Conference on Computer Vision and Pattern Recognition}}.
  \bibinfo{pages}{9664--9674}.
\newblock


\bibitem[Li et~al\mbox{.}(2019)]%
        {li2019qtune}
\bibfield{author}{\bibinfo{person}{Guoliang Li}, \bibinfo{person}{Xuanhe Zhou},
  \bibinfo{person}{Shifu Li}, {and} \bibinfo{person}{Bo Gao}.}
  \bibinfo{year}{2019}\natexlab{}.
\newblock \showarticletitle{Qtune: A query-aware database tuning system with
  deep reinforcement learning}.
\newblock \bibinfo{journal}{\emph{Proceedings of the VLDB Endowment}}
  \bibinfo{volume}{12}, \bibinfo{number}{12} (\bibinfo{year}{2019}),
  \bibinfo{pages}{2118--2130}.
\newblock


\bibitem[Li and Hoiem(2017)]%
        {li2017learning}
\bibfield{author}{\bibinfo{person}{Zhizhong Li} {and} \bibinfo{person}{Derek
  Hoiem}.} \bibinfo{year}{2017}\natexlab{}.
\newblock \showarticletitle{Learning without forgetting}.
\newblock \bibinfo{journal}{\emph{IEEE transactions on pattern analysis and
  machine intelligence}} \bibinfo{volume}{40}, \bibinfo{number}{12}
  (\bibinfo{year}{2017}), \bibinfo{pages}{2935--2947}.
\newblock


\bibitem[Liu et~al\mbox{.}(2018)]%
        {liu2018rotate}
\bibfield{author}{\bibinfo{person}{Xialei Liu}, \bibinfo{person}{Marc Masana},
  \bibinfo{person}{Luis Herranz}, \bibinfo{person}{Joost Van~de Weijer},
  \bibinfo{person}{Antonio~M Lopez}, {and} \bibinfo{person}{Andrew~D
  Bagdanov}.} \bibinfo{year}{2018}\natexlab{}.
\newblock \showarticletitle{Rotate your networks: Better weight consolidation
  and less catastrophic forgetting}. In \bibinfo{booktitle}{\emph{2018 24th
  International Conference on Pattern Recognition (ICPR)}}. IEEE,
  \bibinfo{pages}{2262--2268}.
\newblock


\bibitem[Lopez-Paz and Oquab(2016)]%
        {lopez2016revisiting}
\bibfield{author}{\bibinfo{person}{David Lopez-Paz} {and}
  \bibinfo{person}{Maxime Oquab}.} \bibinfo{year}{2016}\natexlab{}.
\newblock \showarticletitle{Revisiting classifier two-sample tests}.
\newblock \bibinfo{journal}{\emph{arXiv preprint arXiv:1610.06545}}
  (\bibinfo{year}{2016}).
\newblock


\bibitem[Lopez-Paz and Ranzato(2017)]%
        {lopez2017gradient}
\bibfield{author}{\bibinfo{person}{David Lopez-Paz} {and}
  \bibinfo{person}{Marc'Aurelio Ranzato}.} \bibinfo{year}{2017}\natexlab{}.
\newblock \showarticletitle{Gradient episodic memory for continual learning}.
\newblock \bibinfo{journal}{\emph{Advances in neural information processing
  systems}}  \bibinfo{volume}{30} (\bibinfo{year}{2017}),
  \bibinfo{pages}{6467--6476}.
\newblock


\bibitem[Lu et~al\mbox{.}(2018)]%
        {lu2018learning}
\bibfield{author}{\bibinfo{person}{Jie Lu}, \bibinfo{person}{Anjin Liu},
  \bibinfo{person}{Fan Dong}, \bibinfo{person}{Feng Gu}, \bibinfo{person}{Joao
  Gama}, {and} \bibinfo{person}{Guangquan Zhang}.}
  \bibinfo{year}{2018}\natexlab{}.
\newblock \showarticletitle{Learning under concept drift: A review}.
\newblock \bibinfo{journal}{\emph{IEEE Transactions on Knowledge and Data
  Engineering}} \bibinfo{volume}{31}, \bibinfo{number}{12}
  (\bibinfo{year}{2018}), \bibinfo{pages}{2346--2363}.
\newblock


\bibitem[Lu et~al\mbox{.}(2014)]%
        {lu2014concept}
\bibfield{author}{\bibinfo{person}{Ning Lu}, \bibinfo{person}{Guangquan Zhang},
  {and} \bibinfo{person}{Jie Lu}.} \bibinfo{year}{2014}\natexlab{}.
\newblock \showarticletitle{Concept drift detection via competence models}.
\newblock \bibinfo{journal}{\emph{Artificial Intelligence}}
  \bibinfo{volume}{209} (\bibinfo{year}{2014}), \bibinfo{pages}{11--28}.
\newblock


\bibitem[Ma et~al\mbox{.}(2020)]%
        {ma2020active}
\bibfield{author}{\bibinfo{person}{Lin Ma}, \bibinfo{person}{Bailu Ding},
  \bibinfo{person}{Sudipto Das}, {and} \bibinfo{person}{Adith Swaminathan}.}
  \bibinfo{year}{2020}\natexlab{}.
\newblock \showarticletitle{Active learning for ML enhanced database systems}.
  In \bibinfo{booktitle}{\emph{Proceedings of the 2020 ACM SIGMOD International
  Conference on Management of Data}}. \bibinfo{pages}{175--191}.
\newblock


\bibitem[Ma et~al\mbox{.}(2021)]%
        {ma2021learned}
\bibfield{author}{\bibinfo{person}{Qingzhi Ma}, \bibinfo{person}{Ali~Mohammadi
  Shanghooshabad}, \bibinfo{person}{Mehrdad Almasi}, \bibinfo{person}{Meghdad
  Kurmanji}, {and} \bibinfo{person}{Peter Triantafillou}.}
  \bibinfo{year}{2021}\natexlab{}.
\newblock \showarticletitle{Learned Approximate Query Processing: Make it
  Light, Accurate and Fast.}. In \bibinfo{booktitle}{\emph{CIDR}}.
\newblock


\bibitem[Ma and Triantafillou(2019)]%
        {ma2019dbest}
\bibfield{author}{\bibinfo{person}{Qingzhi Ma} {and} \bibinfo{person}{Peter
  Triantafillou}.} \bibinfo{year}{2019}\natexlab{}.
\newblock \showarticletitle{Dbest: Revisiting approximate query processing
  engines with machine learning models}. In
  \bibinfo{booktitle}{\emph{Proceedings of the 2019 International Conference on
  Management of Data}}. \bibinfo{pages}{1553--1570}.
\newblock


\bibitem[Marcus et~al\mbox{.}(2021)]%
        {marcus2021bao}
\bibfield{author}{\bibinfo{person}{Ryan Marcus}, \bibinfo{person}{Parimarjan
  Negi}, \bibinfo{person}{Hongzi Mao}, \bibinfo{person}{Nesime Tatbul},
  \bibinfo{person}{Mohammad Alizadeh}, {and} \bibinfo{person}{Tim Kraska}.}
  \bibinfo{year}{2021}\natexlab{}.
\newblock \showarticletitle{Bao: Making learned query optimization practical}.
  In \bibinfo{booktitle}{\emph{Proceedings of the 2021 International Conference
  on Management of Data}}. \bibinfo{pages}{1275--1288}.
\newblock


\bibitem[Marcus et~al\mbox{.}(2019)]%
        {marcus2019neo}
\bibfield{author}{\bibinfo{person}{Ryan Marcus}, \bibinfo{person}{Parimarjan
  Negi}, \bibinfo{person}{Hongzi Mao}, \bibinfo{person}{Chi Zhang},
  \bibinfo{person}{Mohammad Alizadeh}, \bibinfo{person}{Tim Kraska},
  \bibinfo{person}{Olga Papaemmanouil}, {and} \bibinfo{person}{Nesime Tatbul}.}
  \bibinfo{year}{2019}\natexlab{}.
\newblock \showarticletitle{Neo: A learned query optimizer}.
\newblock \bibinfo{journal}{\emph{arXiv preprint arXiv:1904.03711}}
  (\bibinfo{year}{2019}).
\newblock


\bibitem[McCloskey and Cohen(1989)]%
        {mccloskey1989catastrophic}
\bibfield{author}{\bibinfo{person}{Michael McCloskey} {and}
  \bibinfo{person}{Neal~J Cohen}.} \bibinfo{year}{1989}\natexlab{}.
\newblock \showarticletitle{Catastrophic interference in connectionist
  networks: The sequential learning problem}.
\newblock In \bibinfo{booktitle}{\emph{Psychology of learning and motivation}}.
  Vol.~\bibinfo{volume}{24}. \bibinfo{publisher}{Elsevier},
  \bibinfo{pages}{109--165}.
\newblock


\bibitem[Morningstar et~al\mbox{.}(2021)]%
        {morningstar2021density}
\bibfield{author}{\bibinfo{person}{Warren Morningstar}, \bibinfo{person}{Cusuh
  Ham}, \bibinfo{person}{Andrew Gallagher}, \bibinfo{person}{Balaji
  Lakshminarayanan}, \bibinfo{person}{Alex Alemi}, {and}
  \bibinfo{person}{Joshua Dillon}.} \bibinfo{year}{2021}\natexlab{}.
\newblock \showarticletitle{Density of states estimation for out of
  distribution detection}. In \bibinfo{booktitle}{\emph{International
  Conference on Artificial Intelligence and Statistics}}. PMLR,
  \bibinfo{pages}{3232--3240}.
\newblock


\bibitem[Nalisnick et~al\mbox{.}(2018)]%
        {nalisnick2018deep}
\bibfield{author}{\bibinfo{person}{Eric Nalisnick}, \bibinfo{person}{Akihiro
  Matsukawa}, \bibinfo{person}{Yee~Whye Teh}, \bibinfo{person}{Dilan Gorur},
  {and} \bibinfo{person}{Balaji Lakshminarayanan}.}
  \bibinfo{year}{2018}\natexlab{}.
\newblock \showarticletitle{Do deep generative models know what they don't
  know?}
\newblock \bibinfo{journal}{\emph{arXiv preprint arXiv:1810.09136}}
  (\bibinfo{year}{2018}).
\newblock


\bibitem[Nathan et~al\mbox{.}(2020)]%
        {nathan2020learning}
\bibfield{author}{\bibinfo{person}{Vikram Nathan}, \bibinfo{person}{Jialin
  Ding}, \bibinfo{person}{Mohammad Alizadeh}, {and} \bibinfo{person}{Tim
  Kraska}.} \bibinfo{year}{2020}\natexlab{}.
\newblock \showarticletitle{Learning multi-dimensional indexes}. In
  \bibinfo{booktitle}{\emph{Proceedings of the 2020 ACM SIGMOD International
  Conference on Management of Data}}. \bibinfo{pages}{985--1000}.
\newblock


\bibitem[Nehme et~al\mbox{.}(2009)]%
        {nehme2009self}
\bibfield{author}{\bibinfo{person}{Rimma~V Nehme}, \bibinfo{person}{Elke~A
  Rundensteiner}, {and} \bibinfo{person}{Elisa Bertino}.}
  \bibinfo{year}{2009}\natexlab{}.
\newblock \showarticletitle{Self-tuning query mesh for adaptive multi-route
  query processing}. In \bibinfo{booktitle}{\emph{Proceedings of the 12th
  International Conference on Extending Database Technology: Advances in
  Database Technology}}. \bibinfo{pages}{803--814}.
\newblock


\bibitem[Nguyen et~al\mbox{.}(2015)]%
        {nguyen2015deep}
\bibfield{author}{\bibinfo{person}{Anh Nguyen}, \bibinfo{person}{Jason
  Yosinski}, {and} \bibinfo{person}{Jeff Clune}.}
  \bibinfo{year}{2015}\natexlab{}.
\newblock \showarticletitle{Deep neural networks are easily fooled: High
  confidence predictions for unrecognizable images}. In
  \bibinfo{booktitle}{\emph{Proceedings of the IEEE conference on computer
  vision and pattern recognition}}. \bibinfo{pages}{427--436}.
\newblock


\bibitem[Ostapenko et~al\mbox{.}(2019)]%
        {ostapenko2019learning}
\bibfield{author}{\bibinfo{person}{Oleksiy Ostapenko}, \bibinfo{person}{Mihai
  Puscas}, \bibinfo{person}{Tassilo Klein}, \bibinfo{person}{Patrick
  Jahnichen}, {and} \bibinfo{person}{Moin Nabi}.}
  \bibinfo{year}{2019}\natexlab{}.
\newblock \showarticletitle{Learning to remember: A synaptic plasticity driven
  framework for continual learning}. In \bibinfo{booktitle}{\emph{Proceedings
  of the IEEE/CVF Conference on Computer Vision and Pattern Recognition}}.
  \bibinfo{pages}{11321--11329}.
\newblock


\bibitem[Ovadia et~al\mbox{.}(2019)]%
        {ovadia2019can}
\bibfield{author}{\bibinfo{person}{Yaniv Ovadia}, \bibinfo{person}{Emily
  Fertig}, \bibinfo{person}{Jie Ren}, \bibinfo{person}{Zachary Nado},
  \bibinfo{person}{David Sculley}, \bibinfo{person}{Sebastian Nowozin},
  \bibinfo{person}{Joshua Dillon}, \bibinfo{person}{Balaji Lakshminarayanan},
  {and} \bibinfo{person}{Jasper Snoek}.} \bibinfo{year}{2019}\natexlab{}.
\newblock \showarticletitle{Can you trust your model's uncertainty? evaluating
  predictive uncertainty under dataset shift}.
\newblock \bibinfo{journal}{\emph{Advances in neural information processing
  systems}}  \bibinfo{volume}{32} (\bibinfo{year}{2019}).
\newblock


\bibitem[Park et~al\mbox{.}(2018)]%
        {park2018data}
\bibfield{author}{\bibinfo{person}{Noseong Park}, \bibinfo{person}{Mahmoud
  Mohammadi}, \bibinfo{person}{Kshitij Gorde}, \bibinfo{person}{Sushil
  Jajodia}, \bibinfo{person}{Hongkyu Park}, {and} \bibinfo{person}{Youngmin
  Kim}.} \bibinfo{year}{2018}\natexlab{}.
\newblock \showarticletitle{Data synthesis based on generative adversarial
  networks}.
\newblock \bibinfo{journal}{\emph{arXiv preprint arXiv:1806.03384}}
  (\bibinfo{year}{2018}).
\newblock


\bibitem[Pereira et~al\mbox{.}(2009)]%
        {pereira2009machine}
\bibfield{author}{\bibinfo{person}{Francisco Pereira}, \bibinfo{person}{Tom
  Mitchell}, {and} \bibinfo{person}{Matthew Botvinick}.}
  \bibinfo{year}{2009}\natexlab{}.
\newblock \showarticletitle{Machine learning classifiers and fMRI: a tutorial
  overview}.
\newblock \bibinfo{journal}{\emph{Neuroimage}} \bibinfo{volume}{45},
  \bibinfo{number}{1} (\bibinfo{year}{2009}), \bibinfo{pages}{S199--S209}.
\newblock


\bibitem[Rebuffi et~al\mbox{.}(2017)]%
        {rebuffi2017icarl}
\bibfield{author}{\bibinfo{person}{Sylvestre-Alvise Rebuffi},
  \bibinfo{person}{Alexander Kolesnikov}, \bibinfo{person}{Georg Sperl}, {and}
  \bibinfo{person}{Christoph~H Lampert}.} \bibinfo{year}{2017}\natexlab{}.
\newblock \showarticletitle{icarl: Incremental classifier and representation
  learning}. In \bibinfo{booktitle}{\emph{Proceedings of the IEEE conference on
  Computer Vision and Pattern Recognition}}. \bibinfo{pages}{2001--2010}.
\newblock


\bibitem[Ren et~al\mbox{.}(2019)]%
        {ren2019likelihood}
\bibfield{author}{\bibinfo{person}{Jie Ren}, \bibinfo{person}{Peter~J Liu},
  \bibinfo{person}{Emily Fertig}, \bibinfo{person}{Jasper Snoek},
  \bibinfo{person}{Ryan Poplin}, \bibinfo{person}{Mark Depristo},
  \bibinfo{person}{Joshua Dillon}, {and} \bibinfo{person}{Balaji
  Lakshminarayanan}.} \bibinfo{year}{2019}\natexlab{}.
\newblock \showarticletitle{Likelihood ratios for out-of-distribution
  detection}.
\newblock \bibinfo{journal}{\emph{Advances in Neural Information Processing
  Systems}}  \bibinfo{volume}{32} (\bibinfo{year}{2019}).
\newblock


\bibitem[Ruff et~al\mbox{.}(2021)]%
        {ruff2021unifying}
\bibfield{author}{\bibinfo{person}{Lukas Ruff}, \bibinfo{person}{Jacob~R
  Kauffmann}, \bibinfo{person}{Robert~A Vandermeulen},
  \bibinfo{person}{Gr{\'e}goire Montavon}, \bibinfo{person}{Wojciech Samek},
  \bibinfo{person}{Marius Kloft}, \bibinfo{person}{Thomas~G Dietterich}, {and}
  \bibinfo{person}{Klaus-Robert M{\"u}ller}.} \bibinfo{year}{2021}\natexlab{}.
\newblock \showarticletitle{A unifying review of deep and shallow anomaly
  detection}.
\newblock \bibinfo{journal}{\emph{Proc. IEEE}} (\bibinfo{year}{2021}).
\newblock


\bibitem[Rusu et~al\mbox{.}(2016)]%
        {rusu2016progressive}
\bibfield{author}{\bibinfo{person}{Andrei~A Rusu}, \bibinfo{person}{Neil~C
  Rabinowitz}, \bibinfo{person}{Guillaume Desjardins}, \bibinfo{person}{Hubert
  Soyer}, \bibinfo{person}{James Kirkpatrick}, \bibinfo{person}{Koray
  Kavukcuoglu}, \bibinfo{person}{Razvan Pascanu}, {and} \bibinfo{person}{Raia
  Hadsell}.} \bibinfo{year}{2016}\natexlab{}.
\newblock \showarticletitle{Progressive neural networks}.
\newblock \bibinfo{journal}{\emph{arXiv preprint arXiv:1606.04671}}
  (\bibinfo{year}{2016}).
\newblock


\bibitem[Savva et~al\mbox{.}(2019)]%
        {savva2019aggregate}
\bibfield{author}{\bibinfo{person}{Fotis Savva}, \bibinfo{person}{Christos
  Anagnostopoulos}, {and} \bibinfo{person}{Peter Triantafillou}.}
  \bibinfo{year}{2019}\natexlab{}.
\newblock \showarticletitle{Aggregate query prediction under dynamic
  workloads}. In \bibinfo{booktitle}{\emph{2019 IEEE International Conference
  on Big Data (Big Data)}}. IEEE, \bibinfo{pages}{671--676}.
\newblock


\bibitem[Sekhari et~al\mbox{.}(2021)]%
        {sekhari2021remember}
\bibfield{author}{\bibinfo{person}{Ayush Sekhari}, \bibinfo{person}{Jayadev
  Acharya}, \bibinfo{person}{Gautam Kamath}, {and}
  \bibinfo{person}{Ananda~Theertha Suresh}.} \bibinfo{year}{2021}\natexlab{}.
\newblock \showarticletitle{Remember what you want to forget: Algorithms for
  machine unlearning}.
\newblock \bibinfo{journal}{\emph{Advances in Neural Information Processing
  Systems}}  \bibinfo{volume}{34} (\bibinfo{year}{2021}),
  \bibinfo{pages}{18075--18086}.
\newblock


\bibitem[Shanghooshabad et~al\mbox{.}(2021)]%
        {shanghooshabad2021pgmjoins}
\bibfield{author}{\bibinfo{person}{Ali~Mohammadi Shanghooshabad},
  \bibinfo{person}{Meghdad Kurmanji}, \bibinfo{person}{Qingzhi Ma},
  \bibinfo{person}{Michael Shekelyan}, \bibinfo{person}{Mehrdad Almasi}, {and}
  \bibinfo{person}{Peter Triantafillou}.} \bibinfo{year}{2021}\natexlab{}.
\newblock \showarticletitle{PGMJoins: Random Join Sampling with Graphical
  Models}. In \bibinfo{booktitle}{\emph{Proceedings of the 2021 International
  Conference on Management of Data}}. \bibinfo{pages}{1610--1622}.
\newblock


\bibitem[Siddiqui et~al\mbox{.}(2020)]%
        {siddiqui2020cost}
\bibfield{author}{\bibinfo{person}{Tarique Siddiqui}, \bibinfo{person}{Alekh
  Jindal}, \bibinfo{person}{Shi Qiao}, \bibinfo{person}{Hiren Patel}, {and}
  \bibinfo{person}{Wangchao Le}.} \bibinfo{year}{2020}\natexlab{}.
\newblock \showarticletitle{Cost models for big data query processing:
  Learning, retrofitting, and our findings}. In
  \bibinfo{booktitle}{\emph{Proceedings of the 2020 ACM SIGMOD International
  Conference on Management of Data}}. \bibinfo{pages}{99--113}.
\newblock


\bibitem[Song et~al\mbox{.}(2007)]%
        {song2007statistical}
\bibfield{author}{\bibinfo{person}{Xiuyao Song}, \bibinfo{person}{Mingxi Wu},
  \bibinfo{person}{Christopher Jermaine}, {and} \bibinfo{person}{Sanjay
  Ranka}.} \bibinfo{year}{2007}\natexlab{}.
\newblock \showarticletitle{Statistical change detection for multi-dimensional
  data}. In \bibinfo{booktitle}{\emph{Proceedings of the 13th ACM SIGKDD
  international conference on Knowledge discovery and data mining}}.
  \bibinfo{pages}{667--676}.
\newblock


\bibitem[Thirumuruganathan et~al\mbox{.}(2020)]%
        {thirumuruganathan2020approximate}
\bibfield{author}{\bibinfo{person}{Saravanan Thirumuruganathan},
  \bibinfo{person}{Shohedul Hasan}, \bibinfo{person}{Nick Koudas}, {and}
  \bibinfo{person}{Gautam Das}.} \bibinfo{year}{2020}\natexlab{}.
\newblock \showarticletitle{Approximate query processing for data exploration
  using deep generative models}. In \bibinfo{booktitle}{\emph{2020 IEEE 36th
  International Conference on Data Engineering (ICDE)}}. IEEE,
  \bibinfo{pages}{1309--1320}.
\newblock


\bibitem[Tian et~al\mbox{.}(2020)]%
        {seq-self-distill}
\bibfield{author}{\bibinfo{person}{Yonglong Tian}, \bibinfo{person}{Yue Wang},
  \bibinfo{person}{Dilip Krishnan}, \bibinfo{person}{Joshua~B. Tenenbaum},
  {and} \bibinfo{person}{Phillip Isola}.} \bibinfo{year}{2020}\natexlab{}.
\newblock \showarticletitle{Rethinking Few-Shot Image Classification: a Good
  Embedding Is All You Need?}. In
  \bibinfo{booktitle}{\emph{https://arxiv.org/abs/2003.11539}}.
\newblock


\bibitem[Titsias et~al\mbox{.}(2019)]%
        {titsias2019functional}
\bibfield{author}{\bibinfo{person}{Michalis~K Titsias},
  \bibinfo{person}{Jonathan Schwarz}, \bibinfo{person}{Alexander G de~G
  Matthews}, \bibinfo{person}{Razvan Pascanu}, {and} \bibinfo{person}{Yee~Whye
  Teh}.} \bibinfo{year}{2019}\natexlab{}.
\newblock \showarticletitle{Functional regularisation for continual learning
  with gaussian processes}.
\newblock \bibinfo{journal}{\emph{arXiv preprint arXiv:1901.11356}}
  (\bibinfo{year}{2019}).
\newblock


\bibitem[Van~Aken et~al\mbox{.}(2017)]%
        {van2017automatic}
\bibfield{author}{\bibinfo{person}{Dana Van~Aken}, \bibinfo{person}{Andrew
  Pavlo}, \bibinfo{person}{Geoffrey~J Gordon}, {and} \bibinfo{person}{Bohan
  Zhang}.} \bibinfo{year}{2017}\natexlab{}.
\newblock \showarticletitle{Automatic database management system tuning through
  large-scale machine learning}. In \bibinfo{booktitle}{\emph{Proceedings of
  the 2017 ACM International Conference on Management of Data}}.
  \bibinfo{pages}{1009--1024}.
\newblock


\bibitem[Wang et~al\mbox{.}(2020b)]%
        {wang2020few}
\bibfield{author}{\bibinfo{person}{K Wang}, \bibinfo{person}{Paul Vicol},
  \bibinfo{person}{Eleni Triantafillou}, {and} \bibinfo{person}{Richard
  Zemel}.} \bibinfo{year}{2020}\natexlab{b}.
\newblock \showarticletitle{Few-shot Out-of-Distribution Detection}. In
  \bibinfo{booktitle}{\emph{ICML Workshop on Uncertainty and Robustness in Deep
  Learning}}.
\newblock


\bibitem[Wang et~al\mbox{.}(2020a)]%
        {wang2020we}
\bibfield{author}{\bibinfo{person}{Xiaoying Wang}, \bibinfo{person}{Changbo
  Qu}, \bibinfo{person}{Weiyuan Wu}, \bibinfo{person}{Jiannan Wang}, {and}
  \bibinfo{person}{Qingqing Zhou}.} \bibinfo{year}{2020}\natexlab{a}.
\newblock \showarticletitle{Are We Ready For Learned Cardinality Estimation?}
\newblock \bibinfo{journal}{\emph{arXiv preprint arXiv:2012.06743}}
  (\bibinfo{year}{2020}).
\newblock


\bibitem[Watkins and Mardia(1992)]%
        {watkins1992maximum}
\bibfield{author}{\bibinfo{person}{AJ Watkins} {and} \bibinfo{person}{KV
  Mardia}.} \bibinfo{year}{1992}\natexlab{}.
\newblock \showarticletitle{Maximum likelihood estimation and prediction mean
  square error in the spatial linear model}.
\newblock \bibinfo{journal}{\emph{Journal of Applied Statistics}}
  \bibinfo{volume}{19}, \bibinfo{number}{1} (\bibinfo{year}{1992}),
  \bibinfo{pages}{49--59}.
\newblock


\bibitem[Wilson and Izmailov(2020)]%
        {wilson2020bayesian}
\bibfield{author}{\bibinfo{person}{Andrew~G Wilson} {and}
  \bibinfo{person}{Pavel Izmailov}.} \bibinfo{year}{2020}\natexlab{}.
\newblock \showarticletitle{Bayesian deep learning and a probabilistic
  perspective of generalization}.
\newblock \bibinfo{journal}{\emph{Advances in neural information processing
  systems}}  \bibinfo{volume}{33} (\bibinfo{year}{2020}),
  \bibinfo{pages}{4697--4708}.
\newblock


\bibitem[Wu et~al\mbox{.}(2019)]%
        {wu2019large}
\bibfield{author}{\bibinfo{person}{Yue Wu}, \bibinfo{person}{Yinpeng Chen},
  \bibinfo{person}{Lijuan Wang}, \bibinfo{person}{Yuancheng Ye},
  \bibinfo{person}{Zicheng Liu}, \bibinfo{person}{Yandong Guo}, {and}
  \bibinfo{person}{Yun Fu}.} \bibinfo{year}{2019}\natexlab{}.
\newblock \showarticletitle{Large scale incremental learning}. In
  \bibinfo{booktitle}{\emph{Proceedings of the IEEE/CVF Conference on Computer
  Vision and Pattern Recognition}}. \bibinfo{pages}{374--382}.
\newblock


\bibitem[Wu et~al\mbox{.}(2021)]%
        {wu2021unified}
\bibfield{author}{\bibinfo{person}{Ziniu Wu}, \bibinfo{person}{Peilun Yang},
  \bibinfo{person}{Pei Yu}, \bibinfo{person}{Rong Zhu}, \bibinfo{person}{Yuxing
  Han}, \bibinfo{person}{Yaliang Li}, \bibinfo{person}{Defu Lian},
  \bibinfo{person}{Kai Zeng}, {and} \bibinfo{person}{Jingren Zhou}.}
  \bibinfo{year}{2021}\natexlab{}.
\newblock \showarticletitle{A unified transferable model for ml-enhanced dbms}.
\newblock \bibinfo{journal}{\emph{arXiv preprint arXiv:2105.02418}}
  (\bibinfo{year}{2021}).
\newblock


\bibitem[Xiao et~al\mbox{.}(2020)]%
        {xiao2020likelihood}
\bibfield{author}{\bibinfo{person}{Zhisheng Xiao}, \bibinfo{person}{Qing Yan},
  {and} \bibinfo{person}{Yali Amit}.} \bibinfo{year}{2020}\natexlab{}.
\newblock \showarticletitle{Likelihood regret: An out-of-distribution detection
  score for variational auto-encoder}.
\newblock \bibinfo{journal}{\emph{Advances in neural information processing
  systems}}  \bibinfo{volume}{33} (\bibinfo{year}{2020}),
  \bibinfo{pages}{20685--20696}.
\newblock


\bibitem[Xu et~al\mbox{.}(2019)]%
        {xu2019modeling}
\bibfield{author}{\bibinfo{person}{Lei Xu}, \bibinfo{person}{Maria
  Skoularidou}, \bibinfo{person}{Alfredo Cuesta-Infante}, {and}
  \bibinfo{person}{Kalyan Veeramachaneni}.} \bibinfo{year}{2019}\natexlab{}.
\newblock \showarticletitle{Modeling tabular data using conditional gan}.
\newblock \bibinfo{journal}{\emph{Advances in Neural Information Processing
  Systems}}  \bibinfo{volume}{32} (\bibinfo{year}{2019}).
\newblock


\bibitem[Yang et~al\mbox{.}(2020)]%
        {yang2020neurocard}
\bibfield{author}{\bibinfo{person}{Zongheng Yang}, \bibinfo{person}{Amog
  Kamsetty}, \bibinfo{person}{Sifei Luan}, \bibinfo{person}{Eric Liang},
  \bibinfo{person}{Yan Duan}, \bibinfo{person}{Xi Chen}, {and}
  \bibinfo{person}{Ion Stoica}.} \bibinfo{year}{2020}\natexlab{}.
\newblock \showarticletitle{NeuroCard: one cardinality estimator for all
  tables}.
\newblock \bibinfo{journal}{\emph{arXiv preprint arXiv:2006.08109}}
  (\bibinfo{year}{2020}).
\newblock


\bibitem[Yang et~al\mbox{.}(2019)]%
        {yang2019deep}
\bibfield{author}{\bibinfo{person}{Zongheng Yang}, \bibinfo{person}{Eric
  Liang}, \bibinfo{person}{Amog Kamsetty}, \bibinfo{person}{Chenggang Wu},
  \bibinfo{person}{Yan Duan}, \bibinfo{person}{Xi Chen},
  \bibinfo{person}{Pieter Abbeel}, \bibinfo{person}{Joseph~M Hellerstein},
  \bibinfo{person}{Sanjay Krishnan}, {and} \bibinfo{person}{Ion Stoica}.}
  \bibinfo{year}{2019}\natexlab{}.
\newblock \showarticletitle{Deep unsupervised cardinality estimation}.
\newblock \bibinfo{journal}{\emph{arXiv preprint arXiv:1905.04278}}
  (\bibinfo{year}{2019}).
\newblock


\bibitem[Yim et~al\mbox{.}(2017)]%
        {yim2017gift}
\bibfield{author}{\bibinfo{person}{Junho Yim}, \bibinfo{person}{Donggyu Joo},
  \bibinfo{person}{Jihoon Bae}, {and} \bibinfo{person}{Junmo Kim}.}
  \bibinfo{year}{2017}\natexlab{}.
\newblock \showarticletitle{A gift from knowledge distillation: Fast
  optimization, network minimization and transfer learning}. In
  \bibinfo{booktitle}{\emph{Proceedings of the IEEE Conference on Computer
  Vision and Pattern Recognition}}. \bibinfo{pages}{4133--4141}.
\newblock


\bibitem[Zenke et~al\mbox{.}(2017)]%
        {zenke2017continual}
\bibfield{author}{\bibinfo{person}{Friedemann Zenke}, \bibinfo{person}{Ben
  Poole}, {and} \bibinfo{person}{Surya Ganguli}.}
  \bibinfo{year}{2017}\natexlab{}.
\newblock \showarticletitle{Continual learning through synaptic intelligence}.
  In \bibinfo{booktitle}{\emph{International Conference on Machine Learning}}.
  PMLR, \bibinfo{pages}{3987--3995}.
\newblock


\bibitem[Zhang et~al\mbox{.}(2019)]%
        {zhang2019end}
\bibfield{author}{\bibinfo{person}{Ji Zhang}, \bibinfo{person}{Yu Liu},
  \bibinfo{person}{Ke Zhou}, \bibinfo{person}{Guoliang Li},
  \bibinfo{person}{Zhili Xiao}, \bibinfo{person}{Bin Cheng},
  \bibinfo{person}{Jiashu Xing}, \bibinfo{person}{Yangtao Wang},
  \bibinfo{person}{Tianheng Cheng}, \bibinfo{person}{Li Liu}, {et~al\mbox{.}}}
  \bibinfo{year}{2019}\natexlab{}.
\newblock \showarticletitle{An end-to-end automatic cloud database tuning
  system using deep reinforcement learning}. In
  \bibinfo{booktitle}{\emph{Proceedings of the 2019 International Conference on
  Management of Data}}. \bibinfo{pages}{415--432}.
\newblock


\bibitem[Zhao et~al\mbox{.}(2018)]%
        {zhao2018random}
\bibfield{author}{\bibinfo{person}{Zhuoyue Zhao}, \bibinfo{person}{Robert
  Christensen}, \bibinfo{person}{Feifei Li}, \bibinfo{person}{Xiao Hu}, {and}
  \bibinfo{person}{Ke Yi}.} \bibinfo{year}{2018}\natexlab{}.
\newblock \showarticletitle{Random sampling over joins revisited}. In
  \bibinfo{booktitle}{\emph{Proceedings of the 2018 International Conference on
  Management of Data}}. \bibinfo{pages}{1525--1539}.
\newblock


\bibitem[Zhi~Kang et~al\mbox{.}(2021)]%
        {zhi2021efficient}
\bibfield{author}{\bibinfo{person}{Johan~Kok Zhi~Kang},
  \bibinfo{person}{Sien~Yi Tan}, \bibinfo{person}{Feng Cheng},
  \bibinfo{person}{Shixuan Sun}, {and} \bibinfo{person}{Bingsheng He}.}
  \bibinfo{year}{2021}\natexlab{}.
\newblock \showarticletitle{Efficient Deep Learning Pipelines for Accurate Cost
  Estimations Over Large Scale Query Workload}. In
  \bibinfo{booktitle}{\emph{Proceedings of the 2021 International Conference on
  Management of Data}}. \bibinfo{pages}{1014--1022}.
\newblock


\bibitem[Zhu et~al\mbox{.}(2020)]%
        {zhu2020flat}
\bibfield{author}{\bibinfo{person}{Rong Zhu}, \bibinfo{person}{Ziniu Wu},
  \bibinfo{person}{Yuxing Han}, \bibinfo{person}{Kai Zeng},
  \bibinfo{person}{Andreas Pfadler}, \bibinfo{person}{Zhengping Qian},
  \bibinfo{person}{Jingren Zhou}, {and} \bibinfo{person}{Bin Cui}.}
  \bibinfo{year}{2020}\natexlab{}.
\newblock \showarticletitle{FLAT: Fast, Lightweight and Accurate Method for
  Cardinality Estimation}.
\newblock \bibinfo{journal}{\emph{arXiv preprint arXiv:2011.09022}}
  (\bibinfo{year}{2020}).
\newblock


\bibitem[Zhu et~al\mbox{.}(2019)]%
        {zhu2019novel}
\bibfield{author}{\bibinfo{person}{Yonghua Zhu}, \bibinfo{person}{Weilin
  Zhang}, \bibinfo{person}{Yihai Chen}, {and} \bibinfo{person}{Honghao Gao}.}
  \bibinfo{year}{2019}\natexlab{}.
\newblock \showarticletitle{A novel approach to workload prediction using
  attention-based LSTM encoder-decoder network in cloud environment}.
\newblock \bibinfo{journal}{\emph{EURASIP Journal on Wireless Communications
  and Networking}} \bibinfo{volume}{2019}, \bibinfo{number}{1}
  (\bibinfo{year}{2019}), \bibinfo{pages}{1--18}.
\newblock


\end{thebibliography}

\end{document}